\documentclass[aps,twocolumn,prd,showpacs,floatfix]{revtex4-1}

\usepackage[version=3]{mhchem} 
\usepackage{enumerate}

\usepackage{mathrsfs}
\usepackage{dcolumn}
\usepackage{bm}
\usepackage{amsmath}
\usepackage{amsthm}
\usepackage{amssymb}
\usepackage{graphicx}
\usepackage{footmisc}
\usepackage{nicefrac}
\usepackage{appendix}

\usepackage[normalem]{ulem}
\usepackage[capitalise]{cleveref}

\usepackage[dvipsnames,table,xcdraw]{xcolor}

\newcommand{\eref}[1]{(\ref{#1})}
\newcommand{\eps }{\varepsilon }
\usepackage{epstopdf}

\newcommand{\ben}{\begin{equation}}
\newcommand{\een}{\end{equation}}
\newcommand{\bea}{\begin{eqnarray}}
\newcommand{\eea}{\end{eqnarray}}

\def\br{{\bf r}}
\def\bR{{\bf R}}

\begin{document}

\title{The Coupled-Trajectory Mixed Quantum-Classical Algorithm: A Deconstruction}
\author{Graeme H. Gossel}
\affiliation{Department of Physics and Astronomy, Hunter College and the City University of New York, 695 Park Avenue, New York, New York 10065, USA}
\author{Federica Agostini}
\affiliation{Laboratoire de Chimie Physique, UMR 8000 CNRS/University Paris-Sud, 91405 Orsay, France}
\author{Neepa T. Maitra}
\affiliation{Department of Physics and Astronomy, Hunter College and the City University of New York, 695 Park Avenue, New York, New York 10065, USA} 
\affiliation{The Physics Program and the Chemistry Program of the Graduate Center, City University of New York, 365 Fifth Avenue, New York, USA}

\date{\today}
\begin{abstract}
We analyze a mixed quantum-classical algorithm recently derived from the exact factorization equations [Min, Agostini, Gross, PRL {\bf 115}, 073001 (2015)] to show the role of the different terms in the algorithm in bringing about decoherence and wavepacket branching. The algorithm has the structure of Ehrenfest equations plus a ``coupled-trajectory" term for both the electronic and nuclear equations, and we analyze the relative roles played by the different non-adiabatic terms in these equations, including how they are computed in practise. In particular, we show that while the coupled-trajectory term in the electronic equation is essential in yielding accurate dynamics, that in the nuclear equation has a much smaller effect. A decoherence time is extracted from the electronic equations and compared with that of augmented fewest-switches surface-hopping. 
We revisit a series of non-adiabatic Tully model systems to illustrate our analysis. 

\end{abstract}

\maketitle  
\section{Introduction}
Fascinating phenomena result from the dynamics of correlated electron-ion motion, from the photo-induced processes in photovoltaics~
\cite{PDP09}, to light-induced molecular motors~\cite{Feringa_2017}, cis-trans isomerization~\cite{LM07} ubiquitous in photobiological processes such as vision~\cite{Horst2004}, and photo-dissociation and photo-catalysis in general. When strong laser fields are applied, there is, for example, Coulomb explosion, enhanced ionization, and the prospect of control of chemical reactions via attosecond lasers~\cite{Vrakking_NP2014,Martin2017}. Theoretical and computational modeling of these processes play a crucial role in interpreting, understanding, and predicting the experimental results. To keep the method computationally tractable, typically the nuclei are treated as classical point particles, and most often either the mean-field Ehrenfest scheme is used for the nuclear-electronic feedback~\cite{M64,T98}, or, trajectory surface-hopping~\cite{TP71,T90,T98}. The latter is generally believed to be more accurate, especially since it is able to capture "wavepacket splitting": the fact that after a region of strong coupling between electronic and nuclear degrees of freedom, a nuclear distribution that was initially localized can split into separate parts whose evolution is dictated by different electronic adiabatic states. 
Trajectory surface-hopping has had a tremendous impact in modeling photochemical dynamics~\cite{Subotnik_ARPC2016, Prezhdo_JPCL2016}, since it captures a lot of the correct physics, although it has proved challenging to derive in a consistent way; see Ref.~\cite{LZ17}  for recent progress. 

Recently a novel algorithm was proposed~\cite{MAG15} based on the "exact factorization approach" to coupled dynamics~\cite{AMG10,AMG12}. This coupled-trajectory mixed quantum-classical algorithm (CT-MQC) has been tested on a series of Tully model systems~\cite{AMAG16} as well as on real molecules, and for example, successfully captured photochemical ring-opening and decoherence events in the oxirane molecule~\cite{MATG17}. A crucial element in the algorithm is the coupling of nuclear trajectories, leading to terms that are non-local in the sense that the equations for a given nuclear trajectory and its associated electronic coefficient depend on the positions of neighboring nuclear trajectories. These terms appear alongside terms that are present in conventional mixed quantum-classical treatments, yielding an overall structure of Ehrenfest plus coupled-trajectory (CT) corrections. 

In this paper, we make a detailed study of the roles of the different terms in the CT-MQC equations, discussing how the transfer of electronic population between Born-Oppenheimer (BO) surfaces leads to wavepacket splitting and decoherence. We show that
while the coupled-trajectory term in the electronic equation is crucial to capture both branching and decoherence, the coupled-trajectory term in the nuclear equation is not essential for these effects and often provides only a small quantitative correction. In fact using Ehrenfest forces for the nuclei, coupled to the electronic equation of the CT-MQC algorithm, gives dynamics very close to that of the full algorithm in  a wide range of non-adiabatic situations. Four types of Tully model systems are used to support our analysis of the equations: extended coupling, single-avoided crossing,  double-arch, and a dual avoided-crossing.  
We define a measure of the decoherence time arising from CT-MQC, and compare this with its analog in augmented fewest-switches surface-hopping (A-FSSH)~\cite{SS11,LS12,SOL13}, finding qualitative agreement regarding the onset of
decoherence, but differences in size. 

The paper is organized as follows. The first section reviews the CT-MQC algorithm. The second section analyzes the terms in the CT-MQC equations, elucidating the role of the coupled-trajectory terms in each of the electronic and nuclear equations. The extended-coupling model with a relatively high incoming momentum is used to illustrate our results  in some detail, showing nuclear densities, forces, and electronic coefficients, resolved in time and in space. We then turn to the other three model systems, as well as showing spatially-averaged populations, indicators of decoherence, and branching ratios. This section ends with an exploration of alternative ways to construct the quantum momentum  that enters into the definition of the coupled-trajectory terms. The third section extracts a decoherence time from CT-MQC and compares this with that of A-FSSH. Finally, we present conclusions and an outlook.

\section{The Coupled-Trajectory Mixed Quantum-Classical Algorithm}
In the exact factorization approach, the exact time-dependent molecular Schr\"odinger equation is recast into two coupled equations for the nuclear wavefunction, $\chi(\bR,t)$ and conditional electronic wavefunction, $\Phi_\bR(\br,t)$, whose product yields the exact molecular wavefunction $\Psi(\br,\bR,t) =\chi(\bR,t)\Phi_\bR(\br,t) $ uniquely up to a gauge-like transformation, provided the partial normalization condition, $\int d\br \vert\Phi_\bR(\br,t)\vert^2 = 1$ is satisfied for all nuclear coordinates $\bR$ and time $t$.   We refer the reader to the literature on the exact equations and the properties of the potentials arising in them~\cite{AMG10,AMG12,ACEJ13,AMG13, H75,H81,GG14}, on a factorization where the roles of the electrons and nuclei are reversed~\cite{SAMYG14,KAM15,KARM17}, on mathematical aspects~\cite{JSW15,ML16}, on features of the potentials in exactly-solvable cases~\cite{AASG13, MAKG14, CAG16, FHGS17, RTG16,RPG17, CA17,RPG17,CKOC14, L15,L15b,Curchod_EPJB2018}, on perturbative limits~\cite{SASGV17, SASGV15,EA16,SAG16}, on density-functionalization~\cite{RG16, LRG18}, on nested factorizations~\cite{C15}, on extensions to purely electronic systems~\cite{SG17} and to photonic-electron-nuclear systems~\cite{HARM18}, and on mixed quantum-classical approximations based on the exact factorization formalism~\cite{AMG15,AAG14,AAG14b,AASMMG15,SAMG15, MAG15, AMAG16, SW16,MATG17,HLM18,GF17,Tavernelli_EPJB2018,Gross_bookTDDFT2018}. 

We focus here on  the CT-MQC mixed quantum-classical method~\cite{MAG15,AMAG16,MATG17}, derived from the exact-factorization approach, that accurately captures wavepacket splitting and electronic decoherence. A swarm of nuclear trajectories is evolved via classical dynamics that are coupled to quantum evolution equations for the corresponding electronic wavefunction; the latter equations are for the coefficients of the electronic wavefunction associated with each nuclear trajectory $I$ expanded in the Born-Oppenheimer (BO) basis. 
For the simplest case of one nucleus in one dimension, the equations for this expansion coefficient, $C_{l}^{(I)}(t)$,  and the equation of motion for that trajectory $I$, are
\bea
\dot C_l^{(I)}(t) &=&-\frac{i}{\hbar}\epsilon_{{\rm BO},l}^{(I)}(t) C_{l}^{(I)}(t)- \sum_k C_k^{(I)}(t)\frac{P^{(I)}(t)}{M}d_{lk}^{(I)}(t)\notag{}\\
\label{eq:CT-MQCeqnse}
&  -& \frac{\mathcal{Q}^{(I)}(t)}{\hbar M}\left(\sum_k \vert C_k^{(I)}(t)\vert^2 f_k^{(I)}(t) - f_l^{(I)}(t)\right)C_l^{(I)}(t)\notag{} \\
\eea
and
\bea
\dot{P}^{(I)}(t) &=&M \ddot{R}^{(I)}(t) = -\sum_k \vert C_k^{(I)}(t)\vert^2\nabla_R\epsilon_{{\rm BO},k}^{(I)}(t) \notag{}
\\&-& \sum_{l,k}C_l^{(I)*}(t)C_k^{(I)}(t)\left(\eps_{{\rm BO},k}^{(I)}(t) -\eps_{{\rm BO},l}^{(I)}(t)  \right) d^{(I)}_{lk}(t)\notag{} \\
& -&\left(\frac{2 \mathcal{Q}^{(I)}(t) }{M \hbar} \sum_l \vert C_l^{(I)}(t)\vert^2 f_{l}^{(I)}(t) \right) \notag{} \\
&& \times \left[ \sum_k \vert C_k^{(I)}(t)\vert^2 f_{k}^{(I)}(t)-f^{(I)}_l(t) \right] 
\label{eq:CT-MQCeqnsn}
\eea
where $d_{lk}^{(I)}(t)$ is the first-order non-adiabatic coupling (NAC) between BO states $l$ and $k$ evaluated at coordinate $R = R^{(I)}(t)$, $\epsilon_{{\rm BO},k}^{(I)}(t) = \epsilon_{{\rm BO},k}(R^{(I)}(t))$  is the $k$-th adiabatic electronic surface, and $M$ is the nuclear mass. The quantities particular to the CT-MQC algorithm are the adiabatic force from the $l$-th electronic surface integrated along the path of the $I$-th trajectory, given by
\begin{equation}
f_{l}^{(I)}(t) = - \int^t \nabla_R \epsilon_{BO,l}^{(I)}(t') dt'
\label{eq:intfdiff}
\end{equation}
and the quantum momentum
\begin{equation}
\mathcal{Q}^{(I)}(t) = -\hbar\nabla_{R}\vert\chi^{(I)}(t)\vert/\vert\chi^{(I)}(t)\vert
\label{eq:qmom}
\end{equation}
where $\chi^{(I)}(t) = \chi(R^{(I)}(t))$ is an effective nuclear wavefunction built from the classical nuclear trajectories~\cite{MAG15,AMAG16,MATG17}. 

Thus, the structure of the CT-MQC equations is Ehrenfest plus corrections that couple the classical trajectories via the quantum momentum. It is through these terms that each trajectory and its associated electronic coefficients adjust their dynamics depending on the behavior of the distribution of all other trajectories. 

For the purposes of the analysis below, it will help to have the equation of motion for the electronic populations, $\rho_{ll}^{(I)}(t) =\vert C_{l}^{(I)}(t)\vert^2$, at hand:
\bea\nonumber
\dot \rho_{ll}^{(I)}&=& -\frac{2 P^{(I)}}{M}\sum_k Re(\rho_{lk}^{(I)}) d_{lk}^{(I)} \\
&-&2\frac{\mathcal{Q}^{(I)}}{M}\left(\sum_k \rho_{kk}^{(I)} f_k^{(I)} - f_l^{(I)} \right)\rho^{(I)}_{ll} \;\notag{} \\
\label{eq:pops}
\eea
(assuming the $t$-dependence of all terms as in the previous equations),
for the population associated with trajectory $(I)$. We will refer to $\rho_{ll}^{(I)}(t)$ as the  \textsl{spatially-resolved} or \textsl{trajectory-resolved} population, due its dependence on the trajectory, $I$, and consequently on nuclear positions $R^{(I)}$.
The equation for the off-diagonal term $\rho_{lk}^{(I)}={C_{l}^{(I)}}^{*}C_k^{(I)}$ is
\bea \nonumber
\dot\rho_{lm}^{(I)}&=& i \rho_{lm}^{(I)} (\epsilon_{l}^{(I)} - \epsilon_{m}^{(I)} )-\frac{P^{(I)}}{M}\sum_{k} \left( \rho_{lk}^{(I)} d_{mk}^{(I)} +\rho_{km}^{(I)} d_{lk}^{(I)}\right) \notag{} \\
&-&\frac{\mathcal{Q}^{(I)}}{M}\rho_{lm}^{(I)}\left(2\sum_{k} \rho_{kk}^{(I)} f_{k}^{(I)} - (f_{l}^{(I)} + f_{m}^{(I)}) \right)
\label{eq:offdiag}
\eea
The equations Eqs.~\eref{eq:CT-MQCeqnse}--\eref{eq:CT-MQCeqnsn} are implemented as described in Ref.~\cite{MATG17} and in the Appendix. The quantum momentum is computed by reconstructing the nuclear density as a sum of normalized Gaussians, each centered at a different trajectory. We then approximate the quantum momentum in the form $\mathcal{Q}^{(I)}(t) = \alpha^{(I)}(t)R^{(I)}(t) - R_{0}^{(I)}(t)$. However an adjustment is needed to the algorithm represented by Eqs.~\eref{eq:CT-MQCeqnse}--\eref{eq:CT-MQCeqnsn} because, as is, the equations do not guarantee the physical condition that there is never any  population transfer between electronic states $l$ and $k$ in the absence of a NAC $d^{(I)}_{lk}$ when Eq.~\eref{eq:pops} is summed over all trajectories.
Therefore, the evaluation of the quantum momentum is adjusted in the implementation, in the following way. First, the quantum momentum is decomposed in two-state contributions, in analogy with the NACs that couple always two electronic states, and then a shift $\mathcal{C}_{lk}$ is applied,  $\mathcal{Q}^{(I)} \to \mathcal{Q}^{(I)}_{lk} + \mathcal C_{lk}$,  where ${\mathcal C_{lk}} =R_0^{(I)} - R^0_{lk}$, with $R^0_{lk}$ trajectory-independent, such that
\begin{equation}
\label{eq:NACcondition}
\frac{2}{M}\sum_I \left(\mathcal{Q}^{(I)}_{lk}+\mathcal{C}_{lk}\right)\left( f_k^{(I)} - f_l^{(I)} \right)\rho^{(I)}_{ll}\rho_{kk}^{(I)}=0
\end{equation}
at all times (see supplementary information in Ref.~\cite{MATG17}).  The left-hand-side is
simply the net population change of the entire swarm between states $l$ and $k$, that is enforced to be zero if $d^{(I)}_{lk}=0$ in Eq.~\eref{eq:pops}. Notice that Eq.~\eref{eq:NACcondition} has to be imposed at each time to determine the shift of the quantum momentum, which requires a sum over all trajectories. 

\section{Effect of the coupled-trajectory terms}
We will consider an initial state where a nuclear Gaussian wavepacket, 
\begin{equation}
\chi(R,0) = \frac{1}{( \pi \sigma^2)^{1/4}}\exp\left[-\frac{(R-R_0)^2}{2\sigma^2}+i k_0 (R-R_0)\right]
\end{equation}
 is prepared on a single BO surface, which we denote as $l$, in a region where the NAC is negligible. Initial conditions for the CT-MQC equations~\eref{eq:CT-MQCeqnse}--\eref{eq:CT-MQCeqnsn} are then provided by the Wigner phase-space distribution of $\vert \chi(R,0)\vert^2$, and by $C_m^{(I)}(0) = \delta_{l,m}$ $\forall I$.

\begin{center}
\begin{figure}[h]
\includegraphics[width=.5\textwidth]{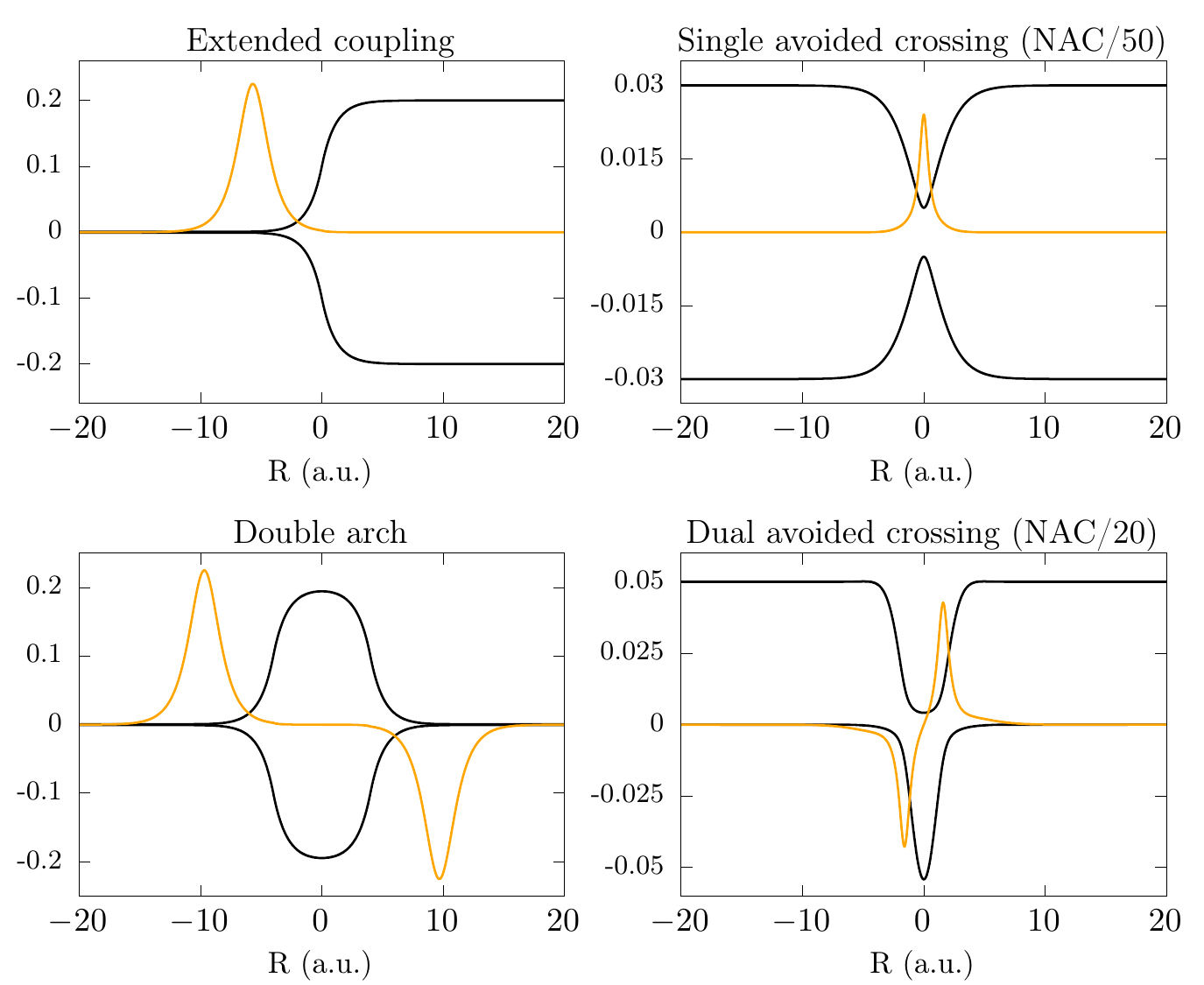}
\caption{
Adiabatic electronic energy surfaces (black) and non-adiabatic couplings (yellow) for four model systems; See Appendix for details of the models.}
\label{fig:BOPES_NAC_fig}
\end{figure}
\end{center}

In what follows, we have chosen a fixed initial width $\sigma = \sqrt{2}$~a.u., and 
selected a few initial momenta $k_0$ that, for all the investigated model systems, provide representative examples. Fig.~\ref{fig:BOPES_NAC_fig} plots the BO surfaces and NACs for the four models we will look at; the details of their parameters are given in the Appendix.

\subsection{Pure Ehrenfest dynamics}
\label{sec:Ehrenfest}
Here we first consider pure Ehrenfest dynamics for the ensemble, that is, we keep only the first two terms on the right-hand-side of each of the equations in Eq.~\eref{eq:CT-MQCeqnse} and Eq.~\eref{eq:CT-MQCeqnsn}, and only the first term in Eq.~\eref{eq:pops} is non-zero. The trajectories in the initial Gaussian swarm begin to evolve following the $l$'th BO surface, with a speed determined by the slope of this surface and the initial momentum of the trajectory~\footnote{In the cases studied here, all surfaces have zero slope in the region where the nuclear wavepacket is initiated, thus its speed and those of the trajectories are simply determined by the initial average momentum.}; only the first term in Eq.~\eref{eq:CT-MQCeqnsn} is non-zero. The associated electronic population does not change initially.  As a nuclear trajectory approaches a region of non-zero NAC between say the $l$'th and $m$'th BO surface, the terms proportional to $d_{lm}$ in Eq.~\eref{eq:CT-MQCeqnse} begin to transfer population from the $l$'th state to the $m$'th state, with the off-diagonal terms $\rho_{lm}$, with $m\neq l$, the first to become non-zero. (This is readily seen, noting that all diagonal elements $d_{kk}=0$, then e.g. for a two-state system:  the first on the right-hand-side of Eq.~\eref{eq:pops}, requires a non-zero $\rho_{l{m\neq l}}^{(I)}$  in order to be active in Eq.~\eref{eq:pops}, while the NAC term  in Eq.~\eref{eq:offdiag} is proportional to $2\rho_{ll}^{(I)} - 1$ so is non-zero even without any population transfer). Consequently, the force driving the trajectory's evolution in Eq.~\eref{eq:CT-MQCeqnsn} begins to mix slopes from both surfaces. 
Population transfer continues until the trajectories leave the interaction region, after which the spatially-resolved populations remain fixed to what they were upon exiting the region: $\rho_{ll}^{(I)}(t)$ becomes constant in time if the NACs between the $l$'th and the other states are zero, since, in Ehrenfest, we have only the first term on the right of Eq.~\eref{eq:pops} and this vanishes if $d_{lm}$ vanishes.
 At this point only the first term in Eq.~\eref{eq:CT-MQCeqnsn} survives, leaving the trajectories to move on a weighted average of the surfaces with the weightings determined by these final electronic populations. Crucially, throughout the evolution, there is no ``communication" between the trajectories and they evolve independently of each other, the only variable determining the dynamics of a given trajectory is its initial energy. Thus, if the spread of energies in the initial wave packet is relatively narrow, the corresponding trajectories, and hence the associated electronic coefficients, follow very similar paths, precluding any splitting of the nuclear wave packet and branching of electronic populations.  Figs.~\ref{fig:T3coeff} and \ref{fig:T7coeff}, which will be extensively discussed in the next section illustrate this for the Ehrenfest dynamics in the extended-coupling and single avoided crossing  model systems: the trajectories and their associated coefficients  in the trailing edge of the wavepacket simply follow those in the leading edge.  The mean-field nature of the dynamics, with neighboring trajectories essentially following each other, does not allow for the possibility of wavepacket splitting.

This lack of communication between the trajectories is the key difference between Ehrenfest and CT-MQC dynamics: in the latter, the evolution of a given trajectory in the swarm depends on where the others are, and the coefficients have a stronger dependence on nuclear positions, thus on the trajectories, as we will show later. It is this trajectory-dependence that enables CT-MQC trajectories to recover splitting of the nuclear wavepacket and branching of the electronic populations.

\subsection{CT-MQC dynamics}
\label{sec:CT-MQCdynamics}
With only one surface $l$ that is initially populated, the dynamics initially follows Ehrenfest: the coupled-trajectory terms are zero in both the electronic and nuclear equations, because the factor $(\sum_m\vert C_m^{(I)}(0)\vert^2 f_m^{(I)}(0) -f_{k}^{(I)}(0))C^{(I)}_k(0)$ can be shown to be zero,  using the following argument. If $l$ is the initially populated state, then $\vert C_m^{(I)}(0)\vert^2=\delta_{lm}$. Then this factor reduces to $(f_l^{(I)}(0) - f_k^{(I)}(0)) C_k^{(I)}(0)$. Again, $C_k^{(I)}(0)\neq 0$, only if $k=l$, which clearly yields $(f_l^{(I)}(0) - f_l^{(I)}(0)) C_l^{(I)}(0) = 0$.  In order for the coupled-trajectory terms to kick in, more than one electronic state must be occupied. Population transfer is first triggered when the trajectory enters a region where a NAC begins to differ from zero, and the coupled-trajectory terms only become effective once there is some fractional population in another BO state because of this factor. In the following, we first discuss the effect of the coupled-trajectory term on the electronic coefficients, before we turn to the nuclear dynamics. We illustrate our analysis in detail first on one of the original Tully models (the extended-coupling model), and then discuss models that cover a range of different non-adiabatic characters.

\subsubsection{Dynamics of the electronic coefficients}
Let us consider a time during a non-adiabatic event  between just two states $l$ and $m$, with $l$ initially fully populated. Then we can simplify Eq.~\eref{eq:pops} to
\bea
\nonumber
\dot{\rho}_{ll}^{(I)} &=& \frac{2}{M}\left [ - P^{(I)} \mathrm{Re}(\rho_{lm}^{(I)}) d_{lm}^{(I)} \right.  \\
\label{eq:2levelrhodot}
&&+ \left. \mathcal{Q}^{(I)}  (f_{l}^{(I)} - f_{m}^{(I)})  (1-\rho_{ll}^{(I)})\rho_{ll}^{(I)}   \right]\\
\nonumber
\dot{\rho}_{mm}^{(I)} &=& -\dot{\rho}_{ll}^{(I)} 
\eea
Henceforth, we will use the symbol $\mathcal Q^{(I)}(t)$ to indicate the quantum momentum between two electronic states, say $l$ and $m$ in Eq.~(\ref{eq:2levelrhodot}), since we will only be concerned with models involving two electronic states, i.e. $\mathcal Q^{(I)} = \mathcal{Q}^{(I)}_{lk}+\mathcal{C}_{lk}$. In more general situations, it is important to indicate the state dependence of the quantum momentum.

At the initial stages of the population transfer, the first term on the right in Eq.~(\ref{eq:2levelrhodot}) dominates: if say a small fraction $\delta$ of trajectories have transferred  to the surface $m$, then the first term goes as $\sqrt{\delta}$  while the second goes as $\delta$. 
However, once the population transfer begins, the second term in Eq.~(\ref{eq:2levelrhodot}) grows  and begins to dominate. 
Unlike the first term,  the second term couples the trajectories so that the change it induces in the population associated with trajectory $I$ depends on the relative position of this trajectory in the swarm through the quantum momentum $\mathcal{Q}^{(I)}$. If the nuclear distribution has retained a  Gaussian shape (albeit wider or narrower, depending on the curvature of the surface $l$), the quantum momentum $\mathcal{Q}^{(I)}$ has a linear shape, positive on the leading edge of the trajectory distribution and negative on the trailing edge, assuming for now $P^{(I)}>0$. This means, that for trajectories for which the integrated force difference $f_{l}^{(I)} - f_{m}^{(I)}$ has the same sign (e.g. those close enough to each other, before significant population transfer), the coupled-trajectory term changes the population associated with trajectories on the leading and trailing edge side in opposite ways, leading to branching of the electronic coefficients. Ultimately this leads to wavepacket splitting as the forces on the trajectories associated with the different surfaces differ (see more shortly in the discussion of $\dot{P}^{(I)}(t)$).

\begin{center}
\begin{figure}[t]
\includegraphics[width=.5\textwidth]{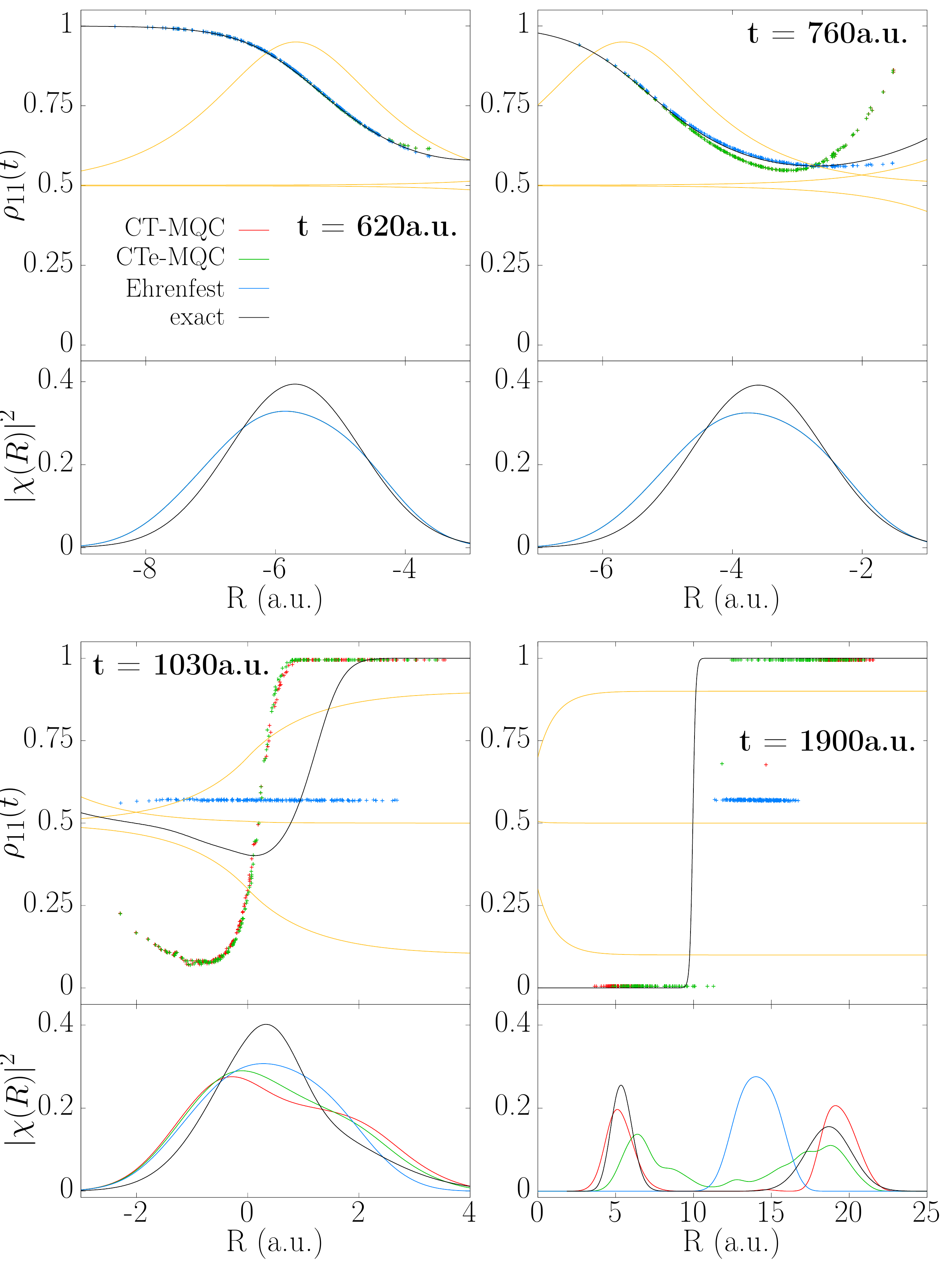}
\caption{Extended-coupling model, with an incoming center momentum of $k_0 = 30$~a.u.: Time snapshots of trajectory-resolved population of the initial electronic state (upper panel) and nuclear densities (lower panel), for full CTMQC (red), CTe-MQC (green), pure Ehrenfest (blue), and exact quantum dynamics (black). Shown in yellow are the two BO surfaces of the model and the NAC.}
\label{fig:T3coeff}
\end{figure}
\end{center}

\begin{center}
\begin{figure}[t]
\includegraphics[width=.5\textwidth]{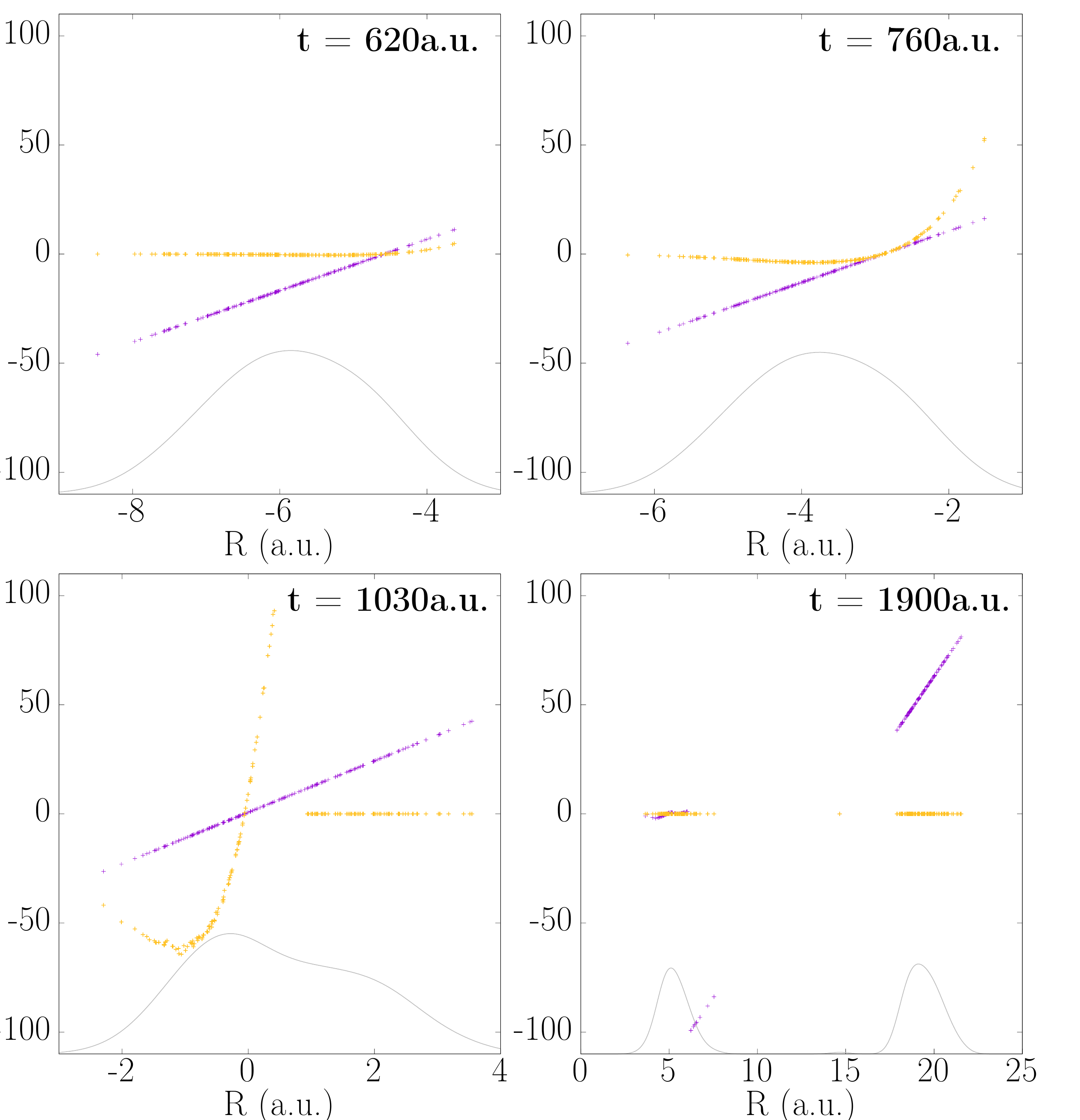}
\caption{Extended-coupling model, with an incoming center momentum of $k_0 = 30$~a.u.: Time snapshots of quantum momentum $Q^{(I)}(t)$ (purple) and the quantum momentum multiplied by the integrated force difference $Q^{(I)}(t)\left( f_{1}^{(I)}(t) - f_{2}^{(I)}(t) \right)$ (yellow) for the full CT-MQC algorithm.  Also shown for reference is the (scaled) CT-MQC nuclear density (grey).}
\label{fig:T3qmom}
\end{figure}
\end{center}

To illustrate this, consider a wavepacket initially located at $R_0 = -15$~a.u., to the left of the avoided crossing of the BO PESs shown for the extended-coupling two-level system in the top left panel of Fig.~\ref{fig:BOPES_NAC_fig}. The dynamics of a wavepacket launched on the lower surface with a relatively high momentum ($k_0 = 30$~a.u.) is  analyzed in Figs.~\ref{fig:T3coeff} and \ref{fig:T3qmom} where we show at several time snapshots:  the spatially-resolved electronic population on the lower surface, $\rho_{11}^{(I)}$ (that on the upper surface is simply $\rho_{22}^{(I)} = 1 - \rho_{11}^{(I)}$ $\forall\,\, I$), the nuclear density reconstructed from the distribution of classical trajectories, the quantum momentum $\mathcal{Q}^{(I)}$, and the quantum momentum multiplied by the integrated force-difference, $\mathcal{Q}^{(I)}  (f_{1}^{(I)} - f_{2}^{(I)})$. The nuclear density at the earlier time snapshots shown are almost identical for all the trajectory methods, and it is important to note that they are in fact much closer to the exact density than indicated by the figure: their deviation in the figure is caused by a numerical smoothing procedure used purely for plotting purposes to obtain a smooth picture of the density from the nuclear histogram of trajectories.
We also note here that when the integrated force difference is computed using Eq.~(\ref{eq:intfdiff}), it is multiplied by a term that sets it to zero if one of the populations approaches within a threshold of 1 (or 0) (see also Appendix): certainly the effect of this term in the equations goes to zero when this happens, and by setting this by hand, we ensure that any small numerical error does not lead to a continuous unphysical action of the quantum momentum.

Tracking the trajectory-resolved electronic populations associated with the trajectories as they enter the NAC region
in Fig.~\ref{fig:T3coeff}, we see that at early times (620~a.u. in the figure and earlier) the CT-MQC population matches that of the Ehrenfest, as argued above, and they match the exact quite well too.
The quantity $\mathcal{Q}^{(I)}  (f_{1}^{(I)} - f_{2}^{(I)})$ is very small at $t=620$~a.u. in Fig.~\eref{fig:T3qmom}, and the quantum momentum does not have much effect on the evolution until there is enough population transferred. When this happens (e.g. time 760 a.u. in Fig.~\eref{fig:T3coeff}), we see that $\mathcal{Q}^{(I)}  (f_{1}^{(I)} - f_{2}^{(I)})$, Fig.~\eref{fig:T3qmom}, is large and positive on the leading trajectories.  Whereas the NAC term induces population transfer to the upper state, the quantum momentum term instead now turns these coefficients on the leading edge of the nuclear distribution back to the lower surface, pushing them 
towards $\rho_{11}=1$, rather than decreasing their values towards zero. 
(Explicitly, the second term in Eq.~(\ref{eq:2levelrhodot}) can be shown to counter the effect of the NAC in the first term: the second term is clearly positive for the trajectories on the leading edge as we just argued, while, the first term is negative, since, with a positive classical momentum, at early times $\dot{\rho}_{lm}\simeq 2 \rho_{ll} d_{lm} P \propto d_{lm}P$ if $\rho_{ll}\sim 1$, so Re$\rho_{lm}^{(I)}>0$). 
 For this model system, we see from the slopes of the BO PES's  that the integrated force difference $f_1^{(I)}- f_2^{(I)}$ is always greater than zero so the coupled-trajectory term in Eq.~\eref{eq:2levelrhodot} is always positive when non-zero, nudging the leading edge trajectories towards their original surface.
 
 On the other hand, for the trailing trajectories, $\mathcal{Q}^{(I)}$ is negative so $\mathcal{Q}^{(I)}  (f_{1}^{(I)} - f_{2}^{(I)})$ is negative (see Fig.~\eref{fig:T3qmom}). When it begins to act, the second term enhances  the population transfer induced by the NAC on the trailing edge trajectories. This behavior persists through the passage through the NAC and a little after, so that the CT-MQC populations form a step-like structure, as shown at time 1030~a.u. in Fig.~\eref{fig:T3coeff}, approximating that of the exact populations, albeit a little too enthusiastically (note that the exact results, i.e., based on the solution of the time-dependent Schr\"odinger equation, are only meaningful/numerically accurate in the region of appreciable nuclear density). In contrast, Ehrenfest dynamics predicts that the spatially-resolved electronic populations, $\rho_{11}^{(I)}$, simply follow each other through as described earlier, roughly averaging over the true branched populations, as clearly shown at times 1030~a.u. and 1900~a.u. in Fig.~\eref{fig:T3coeff}. We again emphasize the key to the branching lies in the quantum momentum through which neighboring trajectories can be associated to electronic populations that evolve quite differently.

As the coefficients associated with the trajectories then return to 1 or 0, the coupled-trajectory term in the equation of motion turns off again due to the $\rho_{11}^{(I)}(1 - \rho_{11}^{(I)})$ factor. The populations then remain fixed until the trajectory might encounter another region of NAC (but this does not happen for this incoming momentum with this model). It is evident in the left lower panel of Fig.~\eref{fig:T3coeff}, at time 1030~a.u., that the effect of the quantum momentum in the electronic evolution equation~\eref{eq:2levelrhodot} goes to zero for the leading and trailing trajectories, i.e., for those that are associated to values of $\rho_{11}^{(I)}$ equal to 0 or 1. The trajectories in the middle still feel the effect of the quantum momentum. At later times, $t=1900$~a.u., the branching is complete, decoherence has completely acted on electronic dynamics, and the effect of the quantum momentum is zero on all trajectories. 

From the nuclear density shown in Figs.~\eref{fig:T3coeff} and~\eref{fig:T3qmom}, we see that at the relatively high incoming momentum chosen here, all trajectories are transmitted through the NAC. Those that get transferred to the upper surface propagate more slowly than those on the lower surface, due to its higher potential energy (see more details shortly). This causes the nuclear wavepacket to split soon after passing through the NAC; we see this just beginning to occur at $t = 1030$~a.u. in both the exact density and even more so in the CT-MQC (consistent with its exaggerated branching), and fully split by the last time-snapshot shown, $t = 1900$~a.u.

 Before discussing the nuclear dynamics further, we note that 
clearly, the argument regarding the branching of the coefficients above does not depend on the nuclear distribution being Gaussian; the same phenomena will occur for any generally localized nuclear distribution.


\subsubsection{Dynamics of the nuclear distribution and the CTe-MQC dynamics}
 We begin by considering  simply the first term on the right-hand-side in the nuclear force of Eq.~(\ref{eq:CT-MQCeqnsn}), that is, the weighted average of the BO surfaces, that we denote via $F^{(I)}_{\rm wBO}(t)$. In general, this is the only term that is non-zero until the region of NAC is reached (in our examples where the potential surfaces are asymptotically flat, it is also zero initially). Once the trajectory $I$ has entered the NAC region,  and a population, 
 $\rho_{mm}^{(I)} = 1 - \rho_{ll}^{(I)}$, has transferred, this term may be written as
\ben
F^{(I)}_{\rm wBO}(t) = -\left((1-\rho_{mm}^{(I)})\nabla\epsilon_{{\rm BO},l}^{(I)} + \rho_{mm}^{(I)} \nabla \epsilon_{{\rm BO}, m}\right)
\label{eq:Pdot1}
\een
Now, from the above discussion, CT-MQC trajectories can be associated with very different populations $\rho_{mm}^{(I)}$,  especially if they are on either side of the maximum of the nuclear distribution. This means that they will weigh the forces from the two surfaces differently. Again, consider the specific example of the extended-coupling model discussed above (upper left panel of Fig.~\ref{fig:BOPES_NAC_fig}) with dynamics presented in Fig.~\ref{fig:T3coeff}. Once the quantum momentum term has come into play, trajectories in the leading edge have a decreasing $\rho_{mm}^{(I)}$, so that the force on them from $F^{(I)}_{\rm wBO}(t)$ tends more and more from the slope of the originally occupied $l$-surface, while those in the trailing edge have an increasing $\rho_{mm}^{(I)}$, so the force from the other surface $m$ begins to dominate. These forces are generally different, so the trajectories begin to move apart, leading eventually to the phenomenon of wavepacket splitting. In the extended-coupling model, the force $F^{(I)}_{\rm wBO}(t)$ causes the trajectories in the leading edge to accelerate for a period of time, while causing those in the trailing edge to slow down. 
 Note that this splitting is achieved with simply $F^{(I)}_{\rm wBO}(t)$; i.e. strikingly, wavepacket splitting occurs even without any term explicitly coupling the trajectories in the nuclear equation. 
The strong spatial dependence of the electronic populations via their dependence on $I$ yields different forces in different portion of $R$-space, beyond mean field, on the corresponding trajectory.
 
To verify this, in Fig.~\ref{fig:T3coeff}  are plotted the electronic populations, and the nuclear distributions, obtained by running simply Ehrenfest dynamics for the nuclei (i.e. including only the first two terms in Eq.~(\ref{eq:CT-MQCeqnsn})) but coupled to the full electronic CT-MQC Eq.~(\ref{eq:CT-MQCeqnse}). We shall denote this evolution scheme as CTe-MQC, to emphasize that the coupled-trajectory term only appears in the electronic equation.  We see that, as predicted, this propagation captures nuclear wavepacket splitting and branching of electronic coefficients, as the nuclear distribution and  spatially-resolved electronic populations are very close to the full CT-MQC algorithm. 

  At the earlier times, $t=620$~a.u. and $t=760$~a.u., all approximations, including Ehrenfest, yield very similar nuclear distributions since the wavepacket is entering the non-adiabatic coupling region and the quantum-momentum term has only recently become non-zero and has not had much effect yet on the nuclear distribution. At later times, $t=1030$~a.u. and $t=1900$~a.u., the difference is more evident. In particular, Ehrenfest cannot reproduce the spatial splitting of the nuclear density, but the trajectories remain localized in an unphysical region of configuration space where the density should be nearly zero ($t=1900$~a.u. of Fig.~\ref{fig:T3coeff}). The two branches of the CT-MQC density appear, at the final time, a bit too far apart if compared to the exact density (see Refs.~\cite{Agostini_EPJB2018, Ciccotti_EPJB2018}), but the spatial splitting is overall correctly captured. Quite remarkably CTe-MQC also captures the splitting in very good agreement with exact results, even though non-zero density is observed where the probability should be close to zero.
  The coupled-trajectory term in the nuclear force in CT-MQC is therefore not completely ineffectual, and causes a "cleaner" wavepacket splitting. 
  
These observations can again be understood by analyzing the structure of the quantum momentum, and in particular its effect in the equation for the classical force, Eq.~\eref{eq:CT-MQCeqnsn}. We define as $\dot{P}^{(I)}_{\mathrm{CT}}(t)  = F^{(I)}_{\mathrm{CT}}(t)$ the coupled-trajectory term in Eq.~\eref{eq:CT-MQCeqnsn} where, for our two-state analysis,
\begin{equation}
F^{(I)}_{\mathrm{CT}}(t) = \frac{2}{M} \mathcal{Q}^{(I)} (f_{l} - f_{m})^2 (1-\rho_{ll})\rho_{ll}
\label{eq:Pdot3}
\end{equation}
The sign of $F^{(I)}_{\mathrm{CT}}(t)$ is determined solely by the sign of the quantum momentum $\mathcal{Q}^{(I)}$. Recalling that this is positive for leading trajectories and negative for trailing ones, and that leading trajectories remain on lower surface, we see that this force accelerates trajectories on the original surface and slows those on the upper surface. Comparing with the weighted-BO force component of Eq.~\eref{eq:Pdot1} shows that these two forces act in the same direction, both working towards the splitting of the nuclear wave packet. Neglecting the coupled-trajectory term in the nuclear force can result in a more diffuse nuclear wavepacket, which is evident in the lower right panel of Fig.~\ref{fig:T3coeff} at time 1900~a.u. where the CTe-MQC density is non-zero, but small, in the region where the quantum-mechanical density is close to zero.  The effect of Eq.~\eref{eq:Pdot3}  can be relatively small however because this force affects relatively few trajectories and it turns off when the population transfer associated with the trajectory is complete.

\begin{center}
\begin{figure}[t]
\includegraphics[width=.5\textwidth]{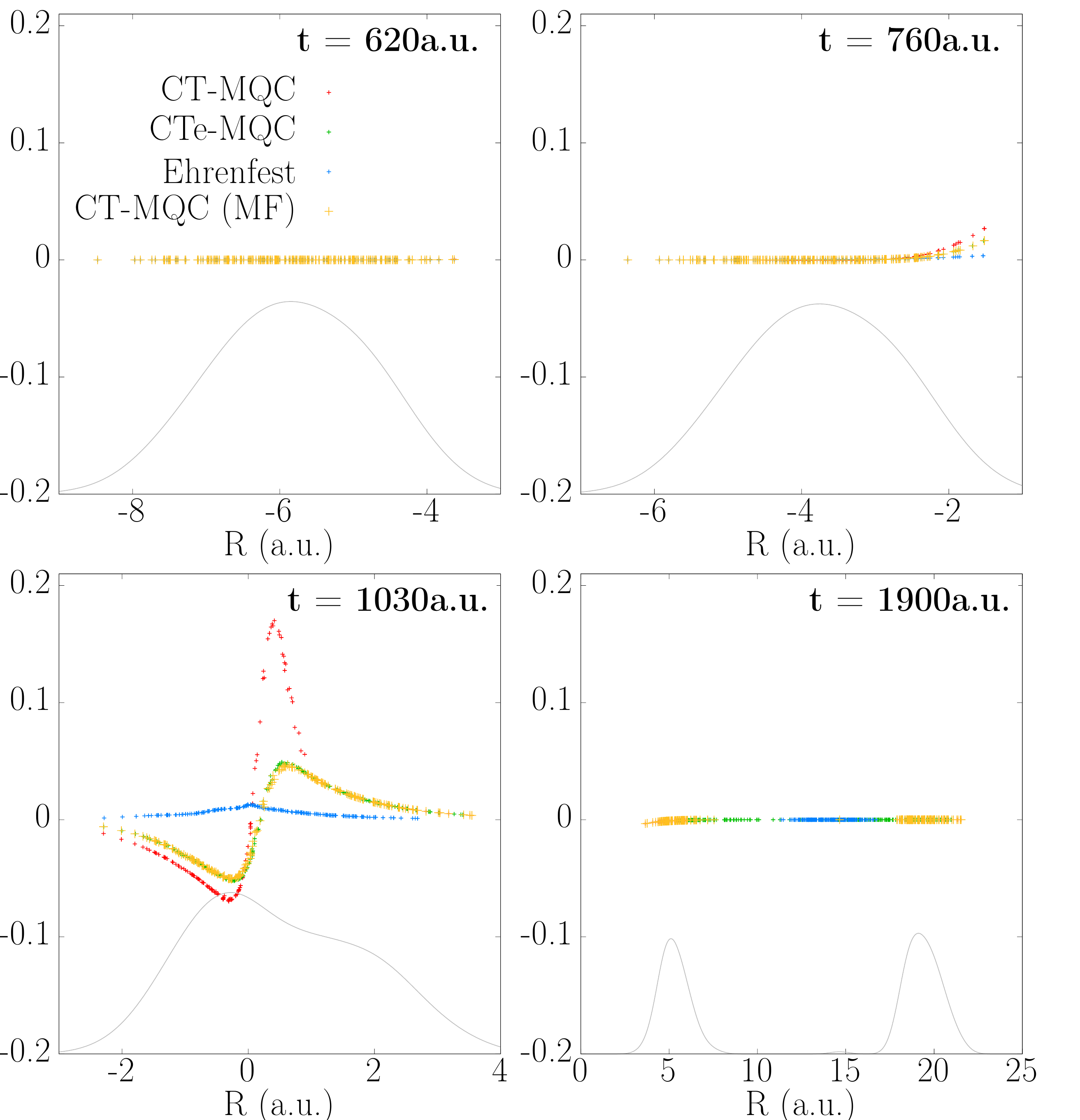}
\caption{Extended-coupling model with initial momentum $k_0 =30$~a.u.:  Time snapshots of the classical forces $\dot{P}^{(I)}(t)$ for CT-MQC (red), CTe-MQC (green), and pure Ehrenfest (light blue), and the mean-field force contribution to the full CT-MQC case indicated by CT-MQC (MF)  (yellow). The (scaled) CT-MQC nuclear density is shown in grey for reference. }
\label{fig:T3pdot}
\end{figure}
\end{center}

For a comparison of the forces, Fig.~\eref{fig:T3pdot} plots snapshots of forces used in Ehrenfest, CT-MQC and CTe-MQC, as well as the mean-field force -- the first two terms in Eq.~\eref{eq:CT-MQCeqnsn} -- for CT-MQC trajectories. As expected, at early times, all contributions to the force are zero (or nearly zero) and the trajectories move only under the effect of the initial momentum. Afterwards, the forces start to have different effects. First, at time $1030$ a.u. we observe that the CTe-MQC dynamics gives rise to an asymmetric classical force of similar magnitude to the full CT-MQC case  and with the same sign, in contrast to the force in purely Ehrenfest dynamics. Second, the effect of $F^{(I)}_{\mathrm{CT}}(t)$ is large only on a narrow subset of the trajectory swarm and only acts for a short period of time. The effect of the coupled-trajectory force term can be estimated from the difference between the full CT-MQC force and the mean-field force term computed for CT-MQC trajectories. Finally, we observe that the total CTe-MQC force and the Ehrenfest-force contribution in CT-MQC  are almost identical even when $F^{(I)}_{\mathrm{CT}}(t)$ term is significant: this result suggests that the addition of this term does not change the dynamics up to this point (time 1030 a.u.); it acts for a short time only after the trajectories have passed through the NAC region.

\subsubsection{Importance of coupling trajectories in the electronic equation}

The analysis above showed how the coupled-trajectory term in the electronic equation leads to branching of the electronic populations, and how this in turn can lead to accurate wavepacket splitting, with even just using a mean-field-like expression for the nuclear force. Conversely, we can ask what happens if we include the coupled-trajectory term only in the nuclear equation, coupled to electronic dynamics determined by Ehrenfest. In this case, where only the first two terms in Eq.~\eref{eq:CT-MQCeqnse} are kept, the coefficients  are directly coupled only by the NAC. This term affects trajectories traveling in the leading and trailing edges in the same way, so neighboring trajectories undergo similar dynamics, and the coefficients cannot branch (see analysis in the Ehrenfest dynamics section)~\footnote{One might imagine the coupled-trajectory term in the nuclear equation can allow the classical momentum $P^{(I)}(t) $ to split due to the structure of $\mathcal{Q}^{(I)}$, which could possibly lead to branching of the electronic populations. However, this did not arise in any of the examples we considered.}. In fact, the splitting of the nuclear wavefunction cannot be captured either, as we will now show.

\begin{center}
\begin{figure}[t]
\includegraphics[width=.5\textwidth]{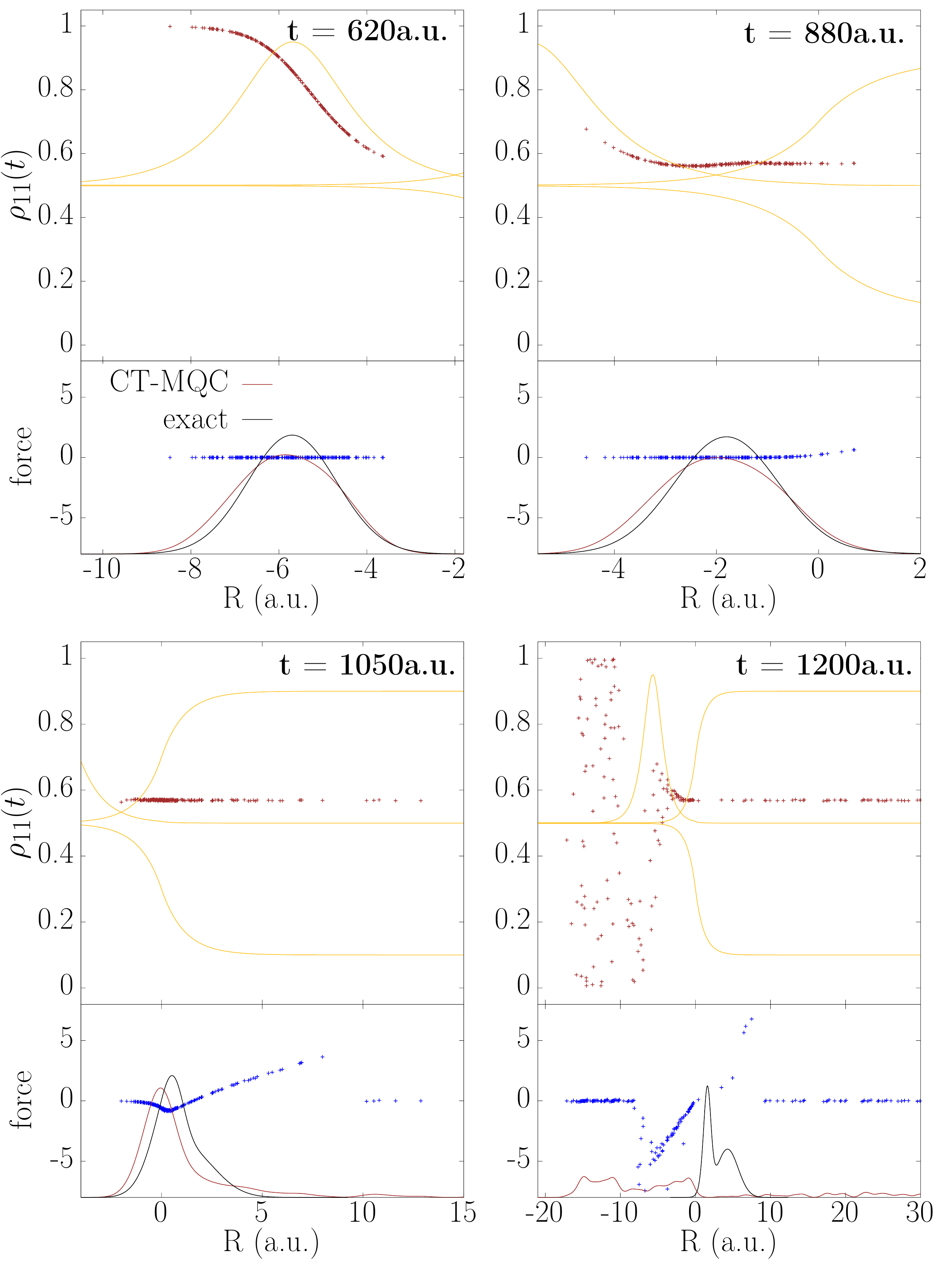}
\caption{Extended-coupling model with incoming momentum $k_0 = 30$~a.u.: Time snapshots of the electronic coefficients (red-brown dots), nuclear densities (red-brown lines; scaled by 25), and total classical forces (blue dots) computed when keeping the coupled-trajectory term only in the nuclear equation (CTn-MQC). In black are shown the exact nuclear densities for reference, and yellow show the BO surfaces and NAC. }
\label{fig:T3_k30_noCTe}
\end{figure}
\end{center}

The effect on the nuclear force component $F^{(I)}_{\rm wBO}(t)$ of keeping only the NAC terms in the electronic equation is that neighboring trajectories feel similar forces, in contrast to what was observed when the coupled-trajectory term is kept in the electronic equations. On the other hand, the force from the coupled-trajectory term $F^{(I)}_{\mathrm{CT}}(t)$ will tend to act in opposite directions for trajectories on either side of the center peak of the nuclear distribution, just as it did in the full CT-MQC algorithm.
However there is a crucial difference with the effect of this term when the coupled-trajectory term is not included in the electronic equations: in the full CT-MQC algorithm,  $F^{(I)}_{\mathrm{CT}}(t)$ tends to zero once full population transfer (of a given trajectory) is complete, due to the population factors collapsing to 1 or 0, but without this collapse, the force continues to remain non-zero. This force is proportional to the integrated force difference which tends to a constant at long times, in contrast to the weighted BO force which is computed at the instantaneous position of the trajectory and which would go to zero if the surfaces asymptotically flatten out. 
So, if the electronic populations never completely branch and decay to either 0 or 1, the force term $F^{(I)}_{\mathrm{CT}}(t)$ may not tend to zero, giving rise to a persistent force that splits the trajectories, leading eventually to unphysical dispersion of the trajectory swarm. Such a situation is depicted in \cref{fig:T3_k30_noCTe} for the extended-coupling case that we have been discussing. This plots the results of propagation when the coupled-trajectory term is kept in the nuclear equation but not in the electronic equation, which we denote as CTn-MQC. 
In particular at time snapshots 1050~a.u. and 1200 a.u. shown, the density in such a dynamics is complete different to the exact one, with much more reflection and much more diffusion. 

Therefore the inclusion of the coupled-trajectory term in the electronic equations is of paramount importance: without it, 
coefficients essentially follow Ehrenfest dynamics with no branching and the nuclear wavepacket dynamics is incorrect, particularly so if the coupled-trajectory term is kept in the nuclear equation while neglected in the electronic. The coupled-trajectory term in the electronic equation brings about branching of coefficients and ensuing wavepacket splitting, and, as a result of this, we will see shortly that measures of decoherence are also very well-captured.

\subsection{Illustrations on other model systems}
Above, we demonstrated our analysis on the Tully extended-coupling model of the top left panel of Fig.~\ref{fig:BOPES_NAC_fig} with an incoming momentum high enough that essentially all the wavepacket gets transmitted. Although the essential points of the analysis go through for a range of different momenta and different models, some details are more complicated because of two main factors. First, the extended-coupling model has a special property that the integrated force difference always has the same sign, which makes the analysis  a little simpler than otherwise, since the sign of the coupled-trajectory terms is then entirely determined by the quantum momentum. Typically, for example in a single avoided crossing, this integrated force has different signs in different parts of the nuclear distribution. This change in sign makes the analysis of the effect of the coupled-trajectory terms in the equations slightly more complicated, although  the essential concepts remain the same. 
Second, the shifts introduced by imposing the condition, Eq. (7), can  slightly modify the analysis of the effect of the coupled-trajectory term given in the previous section. 
Depending on the system, this condition can end up giving a large shift to $\mathcal{Q}^{(I)}$, that significantly influences the population transfer and the force on the nuclear trajectory. In fact, such a shift is evident even in the extended-coupling case studied in the previous sections; consider for instance times 620~a.u. and 760 a.u. in Fig.~\ref{fig:T3qmom}. Here the imposition of the condition Eq.~(\ref{eq:NACcondition}) has caused a shift downward in the quantum momentum to what it would have been with the original definition of Eq.~(\ref{eq:qmom}). With the original definition, the quantum momentum would go through zero near the maximum of the nuclear distribution. In particular the sign of the coupled-trajectory terms in the evolution equations are reversed from what they would have been with the original definition, for some of the trajectories to the near right of the density-maximum. 
The essential arguments given there still hold for most trajectories, certainly on the left and to the truly leading edge, and so the analysis given earlier essentially still holds. 

\begin{center}
\begin{figure}[t]
\includegraphics[width=.5\textwidth]{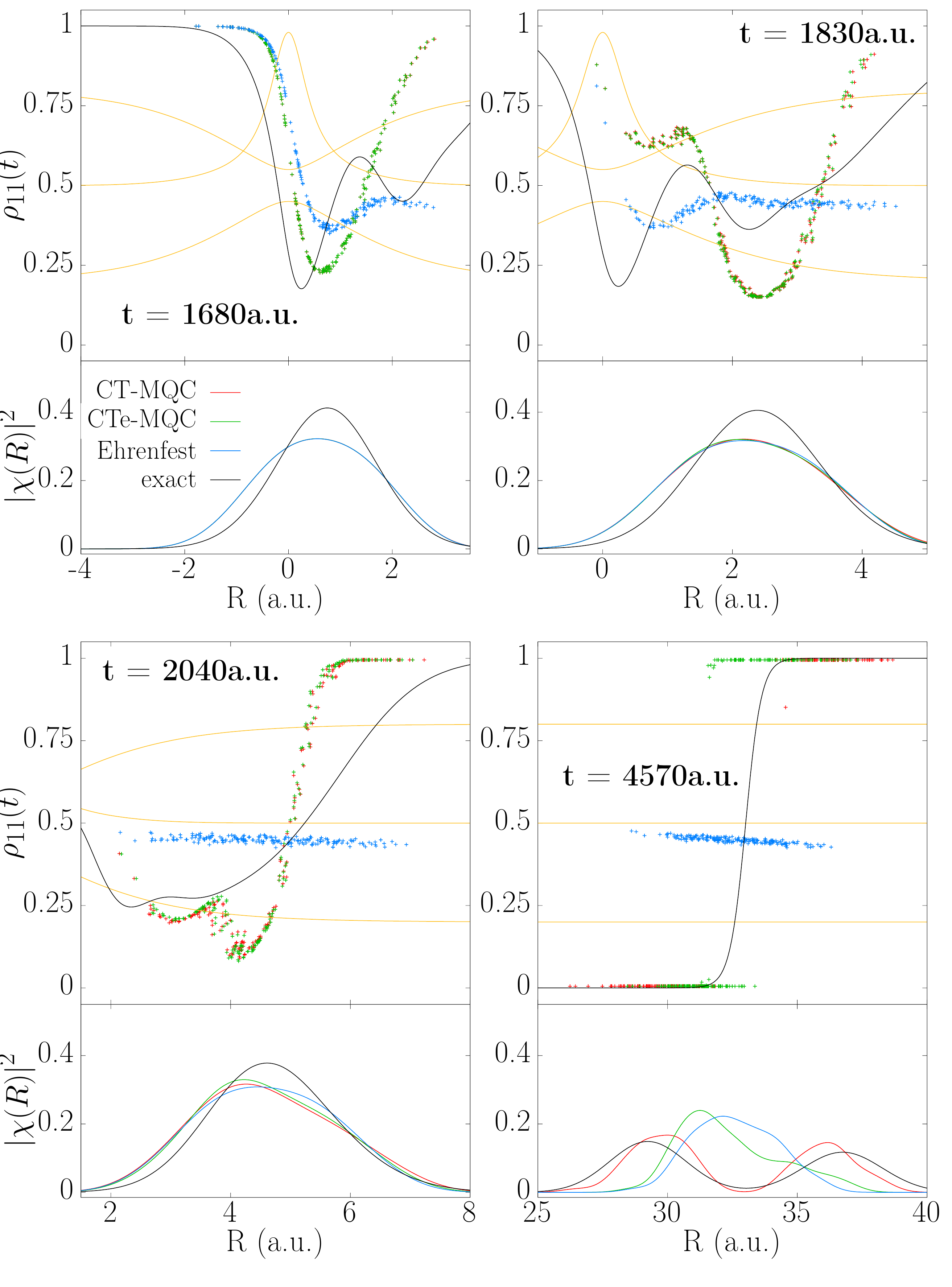}
\caption{Single avoided crossing model with incoming momentum $k_0 = 25$ a.u.: Time snapshots of the trajectory-resolved populations of the initial electronic state (upper panel) and nuclear densities (lower panel) for full CTMQC (red), CTe-MQC (green), pure Ehrenfest (blue), and exact quantum dynamics (black). Shown in yellow are the two BO surfaces of the model and the NAC.}
\label{fig:T7coeff}
\end{figure}
\end{center}

\begin{center}
\begin{figure}[t]
\includegraphics[width=.5\textwidth]{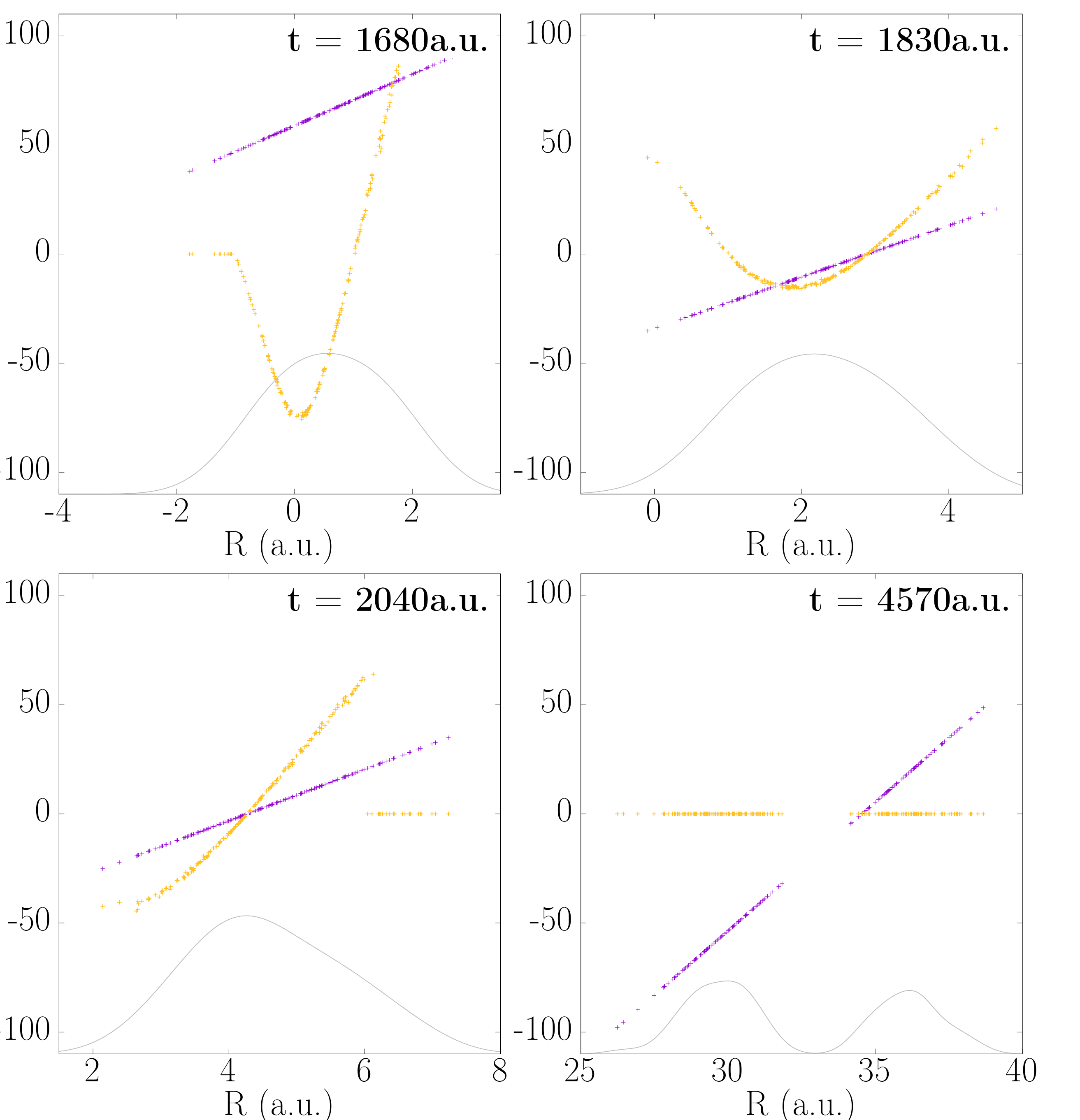}
\caption{Single avoided crossing model with incoming momentum $k_0 = 25$~a.u.: Time snapshots of the quantum momentum $Q^{(I)}(t)$ (purple) and the quantum momentum times the integrated force difference, $Q^{(I)}(t)\left( f_{1}^{(I)}(t) - f_{2}^{(I)}(t) \right)$ (yellow) for the full CT-MQC case. Also shown for reference is the (scaled) CT-MQC nuclear density (grey).}
\label{fig:T7qmom}
\end{figure}
\end{center}

Figs.~\ref{fig:T7coeff} and \ref{fig:T7qmom} show plots analogous to those in 
the previous section for a model system of the single avoided crossing shown on the top right panel of Fig.~\ref{fig:BOPES_NAC_fig}, with an initial momentum of $k_0 = 25$~a.u. In particular, notice that once again the result of CTe-MQC propagation gives results very close to that of the full CT-MQC algorithm for the spatially-resolved coefficients and nuclear distributions for reasons given in the previous section. 

Tracking the trajectories soon after they have entered the NAC region, again the quantum momentum behaves linearly, beginning to kick the leading edge trajectories back towards the initial surface, while enhancing transfer of the trailing edge, separating away from the mean-field Ehrenfest swarm.  But this behavior does not persist for long: the integrated force-difference changes sign soon after the trajectory passes the peak of the NAC,  which leads to some "folds" in the structure of the coefficients. This is particularly evident in the exact quantum coefficients shown  at times 1680~a.u., 1830~a.u. and 2040~a.u. in Fig.~\ref{fig:T7coeff}, and also sometimes captured in the trajectory-resolved coefficients, although, when it is, the CT-MQC algorithm tends to exaggerate the depths of these folds. The imposition of condition Eq.~\eref{eq:NACcondition} apparently leads to large shifts in $\mathcal{Q}^{(I)}$ away from being simply the logarithmic  gradient of the nuclear density, even as the system passes through the NAC (top left panel in Fig.~\ref{fig:T7qmom}). 
In this case, this appears to have the effect of ``smoothing out" the folds that are seen in the exact coefficients: despite the integrated force changing sign through the nuclear distribution, the shift in the quantum momentum that it is multiplied by alters the sign of the term in Eq.~\eref{eq:2levelrhodot} than what it would have been without this shift. For example, both top panels of  Fig.~\ref{fig:T7qmom} would have $\mathcal Q^{(I)}(f_l^{(I)}-f_m^{(I)})$ change sign twice within the swarm, were it not for the shift in $\mathcal Q^{(I)}$ in the top left panel. The effect of this is evident in  Fig.~\ref{fig:T7coeff}, where noticable folds on the coefficients only appear at time $1810$ a.u. once multiple sign changes appear in  $\mathcal Q^{(I)}(f_l^{(I)}-f_m^{(I)})$. At later times this expression resorts to only having a single sign change and the folds begin to be suppressed.
 Once the trajectory leaves the NAC region, the structure of the coefficients simplifies again, and the coupled-trajectory term turns off again once the population transfer is achieved, as described in the section on CT-MQC dynamics. 
 
\begin{center}
\begin{figure}[t]
\includegraphics[width=.5\textwidth]{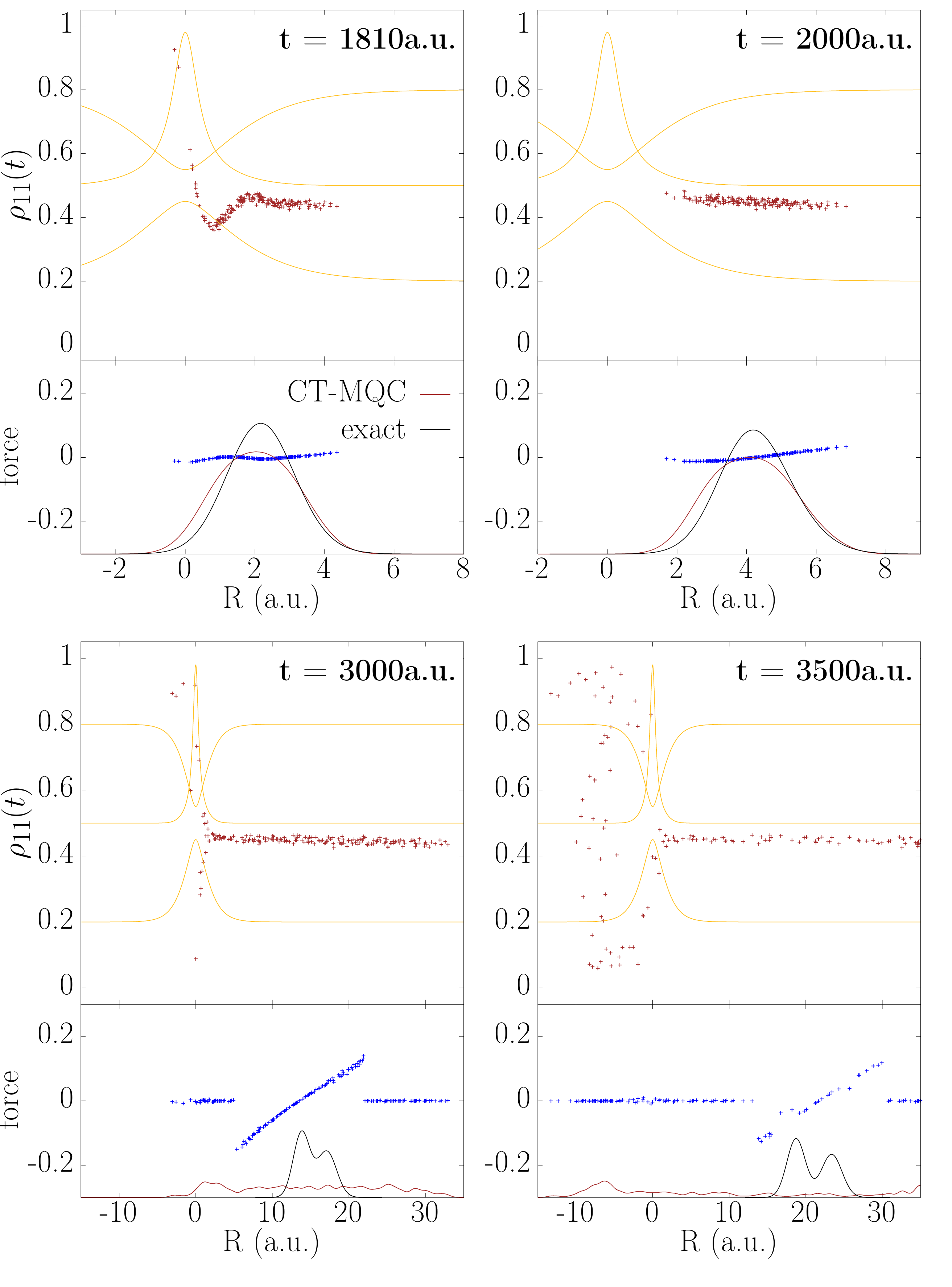}
\caption{Single avoided crossing model with incoming momentum $k_0 = 25$.: Time snapshots of the  electronic coefficients (yellow), nuclear densities (grey; to scale), and total classical forces (blue) computed when keeping the coupled-trajectory term only in the nuclear equation (CTe-MQC). }
\label{fig:T7_k15_noCTe}
\end{figure}
\end{center}

Again, neglecting the coupled-trajectory term in the electronic equation while keeping it in the nuclear equation does not work very well, as shown in Fig.~\ref{fig:T7_k15_noCTe}  for the reasons explained earlier.

\subsubsection{Trajectory-averaged populations, coherences, and branching ratios}

We next define trajectory-averaged populations and indicator of coherences, as
 \begin{align}
\rho_{ll}(t) &=  \frac{1}{N}\sum_{I} \rho_{ll}^{(I)}(t), \notag{} \\
\left| \rho_{lm}(t)\right|^2 &=  \frac{1}{N}\sum_{I} \rho_{ll}^{(I)}(t)\rho_{mm}^{(I)}(t), \notag{} \\
\label{eq:CT_int_defs}
 \end{align}
with $N$ the number of trajectories. The corresponding quantum mechanical expressions are
 \begin{equation}
 \rho_{ll}(t)  = \int |\chi(R,t)|^2 \rho_{ll}(R,t) dR
 \end{equation} 
and
 \begin{equation}
 |\rho_{lm}(t)|^2  = \int |\chi(R,t)|^2 \rho_{ll}(R,t)\rho_{mm}(R,t) dR.
 \end{equation}

\begin{center}
\begin{figure}[h]
\includegraphics[width=.5\textwidth]{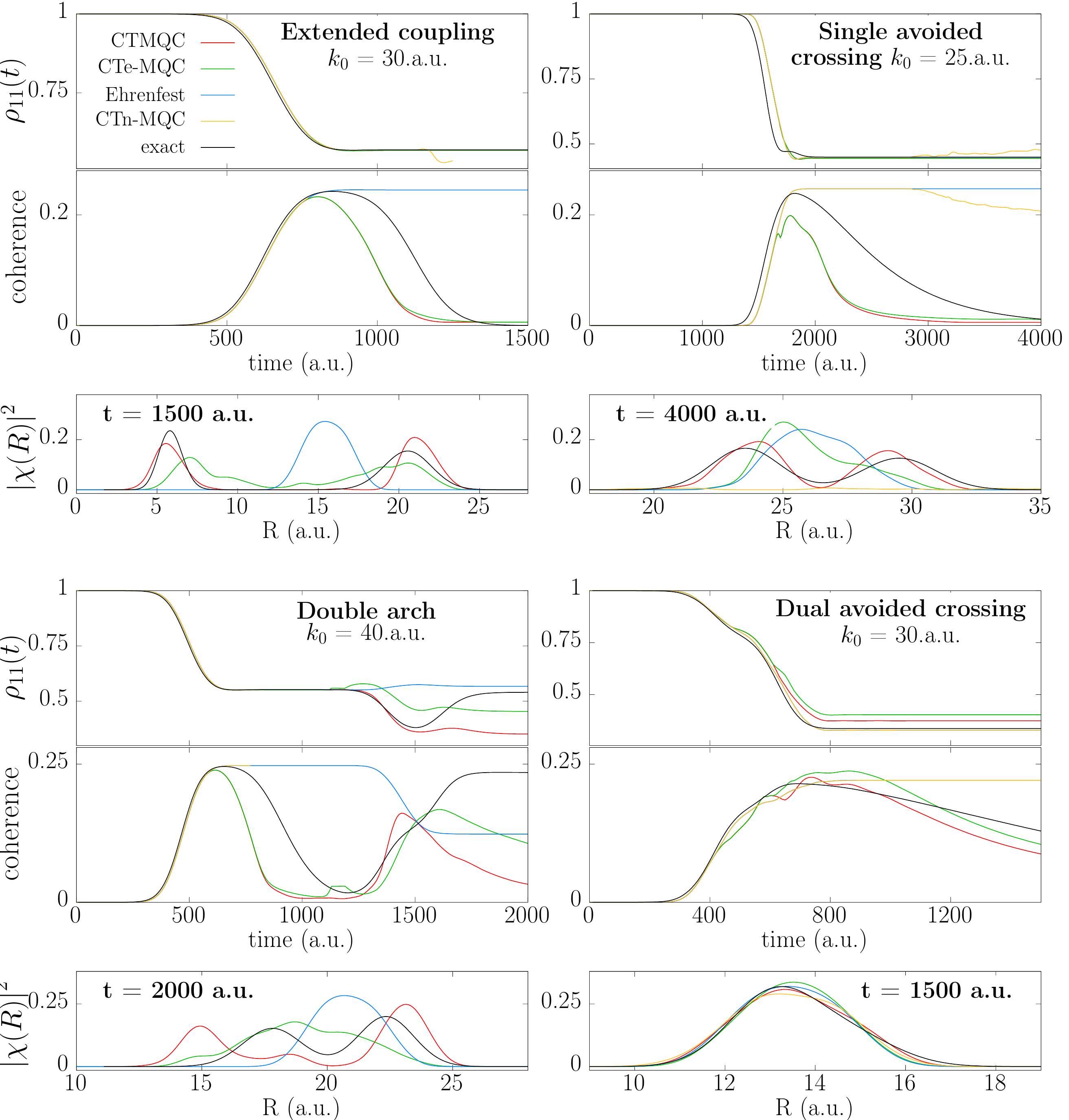}
\caption{Spatially integrated populations (upper panels) and  indicators of coherence (middle panels) given by Eq.~\eref{eq:CT_int_defs} with the nuclear wavepacket having high initial momenta. The bottom panels show the nuclear densities at the final time point. Except for the right hand-side cases, the CTn-MQC simulation fails before this final time point, and so its density is not shown.}
\label{fig:highk_coh_pop}
\end{figure}
\end{center}

\begin{center}
\begin{figure}[h]
\includegraphics[width=.5\textwidth]{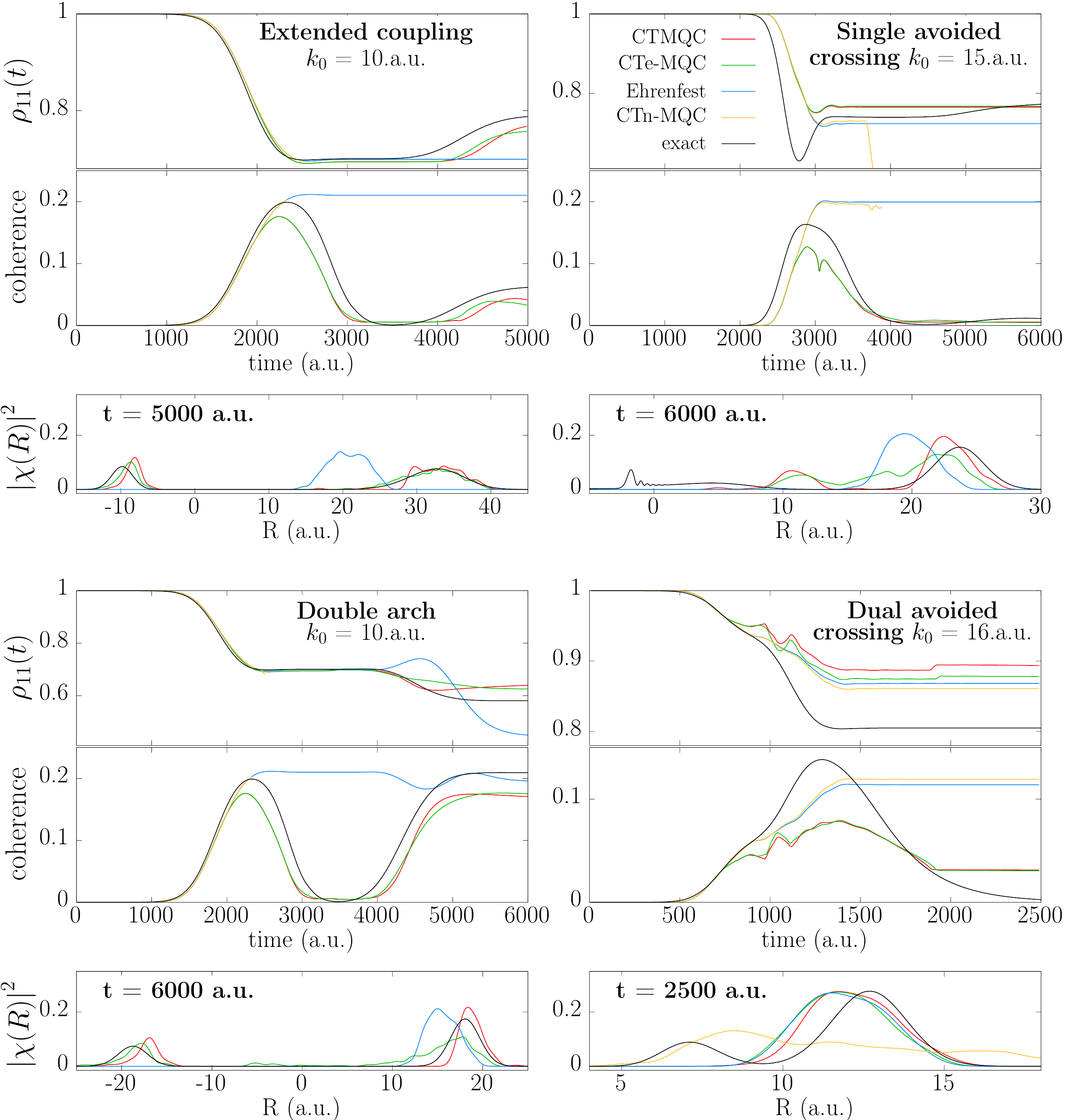}
\caption{Spatially integrated populations (upper panels) and indictors of coherences (middle panels) given by Eq.~\eref{eq:CT_int_defs} with the nuclear wavepacket having low initial momenta. The bottom panels show the nuclear densities at the final time point. Except for the bottom right case, the CTn-MQC simulation fails before this final time point, and so its density is not shown. }
\label{fig:lowk_coh_pop}
\end{figure}
\end{center}

The time trace of the electronic populations and of the indicators of coherence is shown in Figs.~\ref{fig:highk_coh_pop} and~\ref{fig:lowk_coh_pop} for all the models in Fig.~\ref{fig:BOPES_NAC_fig} computed using the exact quantum dynamics, CT-MQC, CTe-MQC, Ehrenfest, and  CTn-MQC. The non-adiabatic character of the dynamics depends on the speed of the initial swarm of trajectories (that is, of the initial nuclear wavepacket). Lower incoming momentum results in less population transfer, meaning that the process is more adiabatic (Fig.~\ref{fig:lowk_coh_pop}), whereas for higher initial momentum the dynamics is more non-adiabatic and there is larger population transfer between the electronic states (Fig.~\ref{fig:highk_coh_pop}). 

We notice that all of the methods do very well for the trajectory-averaged populations in all cases, and especially when only one interaction region has been encountered. A second interaction region is encountered by the exact, CT-MQC, and CTe-MQC calculations in the low-momentum extended-coupling case, the double-arch, and the dual avoided crossing. These trajectory methods capture the branching quite well but not perfectly, as can be seen by the indicator of coherence, which can lead to some discrepancy when a second region is encountered. 

In all cases the CTe-MQC results are very close to the CT-MQC results. In particular, both the full CT-MQC and CTe-MQC capture decoherence in both cases, as the indicator of coherence decays in all situations as expected from the quantum-mechanical calculations. However, such decay seems to occur over shorter time than the exact. Pure Ehrenfest dynamics fails in correctly predicting this decay, because the electronic Ehrenfest equation only contains an oscillatory contribution  in the region where the NACs are zero (see for instance the first two terms of Eq.~\eref{eq:offdiag} representing the electronic Ehrenfest equation for the coherences). The key element to achieve this decaying behavior of the indicator of coherence
is the coupling of the trajectories in the electronic equation. This coupling results in the spatially-resolved coefficients becoming step-like after the interaction, as we have seen in previous sections, giving different weighted averages in the Ehrenfest-like force component that dictates the trajectory evolution. This component alone causes the nuclear trajectories to move apart, each carrying with it its own coefficient, that collapses to a single BO state.
On the other hand, Ehrenfest trajectory-averaged populations are quite accurate, as are the CT-MQC and CTe-MQC ones, even though the Ehrenfest spatially-resolved populations (Fig.~\ref{fig:T3coeff}) are not.


\begin{center}
\begin{figure}[t]
\includegraphics[width=.5\textwidth]{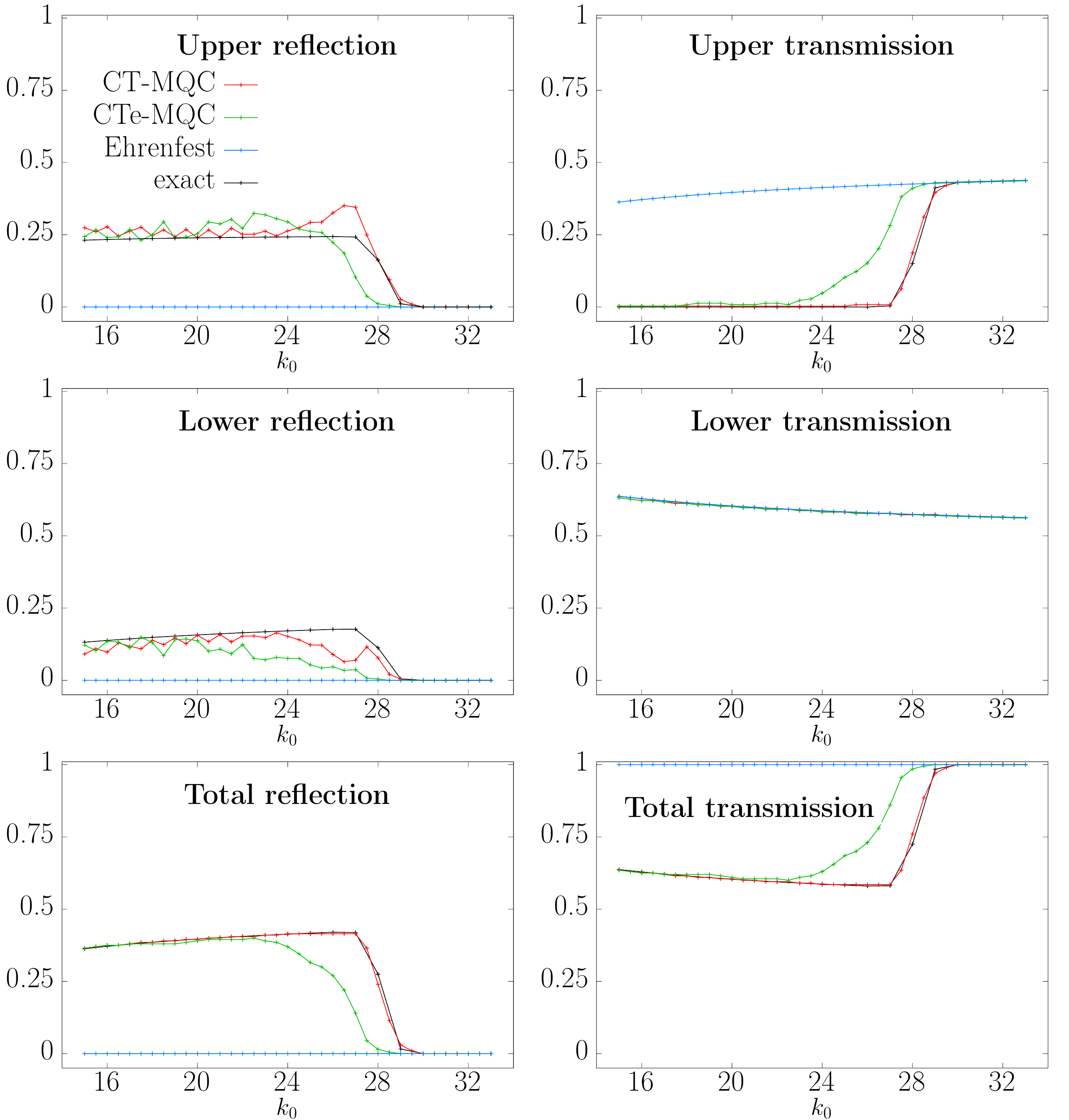}
\caption{Extended-coupling model: branching ratios for CT-MQC (red), CTe-MQC (green), pure Ehrenfest (blue), and exact quantum dynamics (black).}
\label{fig:T3_branching}
\end{figure}
\end{center}

Fig.~\ref{fig:T3_branching} plots branching ratios for the extended-coupling case over a range of initial momenta. Although the total population reflected is well-predicted in the extended-coupling case, there is an overestimate by CT-MQC on the population reflected on the upper surface that cancels its underestimate of the population reflected on the lower surface (compare the left hand figures in Fig.~\ref{fig:T3_branching}). 
Fig.~\ref{fig:T3_branching} also shows that CTe-MQC turns on upper transmission while turning off upper and lower reflection at a lower value of  $k_0$ than that in the exact and full CT-MQC calculations. This is a result of the missing term, Eq.~\eref{eq:Pdot3}, in the equation for the force in CTe-MQC, which as we discussed in the previous sections, leads to the CTe-MQC nuclear distribution being less cleanly split than the full CT-MQC distribution; the trailing edge trajectories feel an additional force to the left from this term while the leading edge trajectories feel an additional force to the right. The wavepacket has momentum already moving to the right,  so when this term is neglected, there are less trajectories reflected and more transmitted in the long-time limit. This explains the underestimation of the reflection and overestimation of the transmission of the CTe-MQC results seen in Fig.~\ref{fig:T3_branching} in the threshold region. 
As the force Eq.~\eref{eq:Pdot3} operates largely during the time from when the nuclear distribution has begun to split to when it has separated into two pieces, the $k_0$ values right around the threshold of the transmission/reflection turn on/off are particularly sensitive to it. 

In some cases, not shown here, branching ratios can discern discrepancies that are missed by the trajectory-averaged populations, while
in other cases (also not shown here) the time-resolved averaged populations are not predicted as good as their final values, which is important to bear in mind when evaluating a method based on branching ratios that are sensitive only to the long-time limit.
Still, both branching ratios and the integrated populations or coherences are averaged quantities, therefore some details of the dynamics are inevitably washed out summing over large number of trajectories. Analysis of spatially-resolved quantities, as $\rho_{ll}^{(I)}(t)$, instead, provides detailed information that helps understanding weaknesses and strengths of the approximations.


\section{Exploring an alternative to imposing Eq.~\eref{eq:NACcondition}}
\label{sec:qmom}
As we have seen in the examples above, the condition Eq.~(\ref{eq:NACcondition}), that ensures no net, i.e. trajectory-averaged, population transfer in the absence of a NAC, can affect the dynamics even when trajectories are in a NAC region.  Here we explore an alternative construction of $\mathcal{Q}^{(I)}$, that preserves the original definition of the quantum momentum as the logarithmic derivative of the nuclear distribution in regions where the NAC is above some threshold value, say $\Delta$, without applying the shift. For trajectories that are in a region such that $\left|d_{lm}\right|<\Delta$, we apply the shift determined by the condition Eq.~\eref{eq:NACcondition} evaluated using only trajectories that are in this region. 

\begin{center}
\begin{figure}[h]
\includegraphics[width=.5\textwidth]{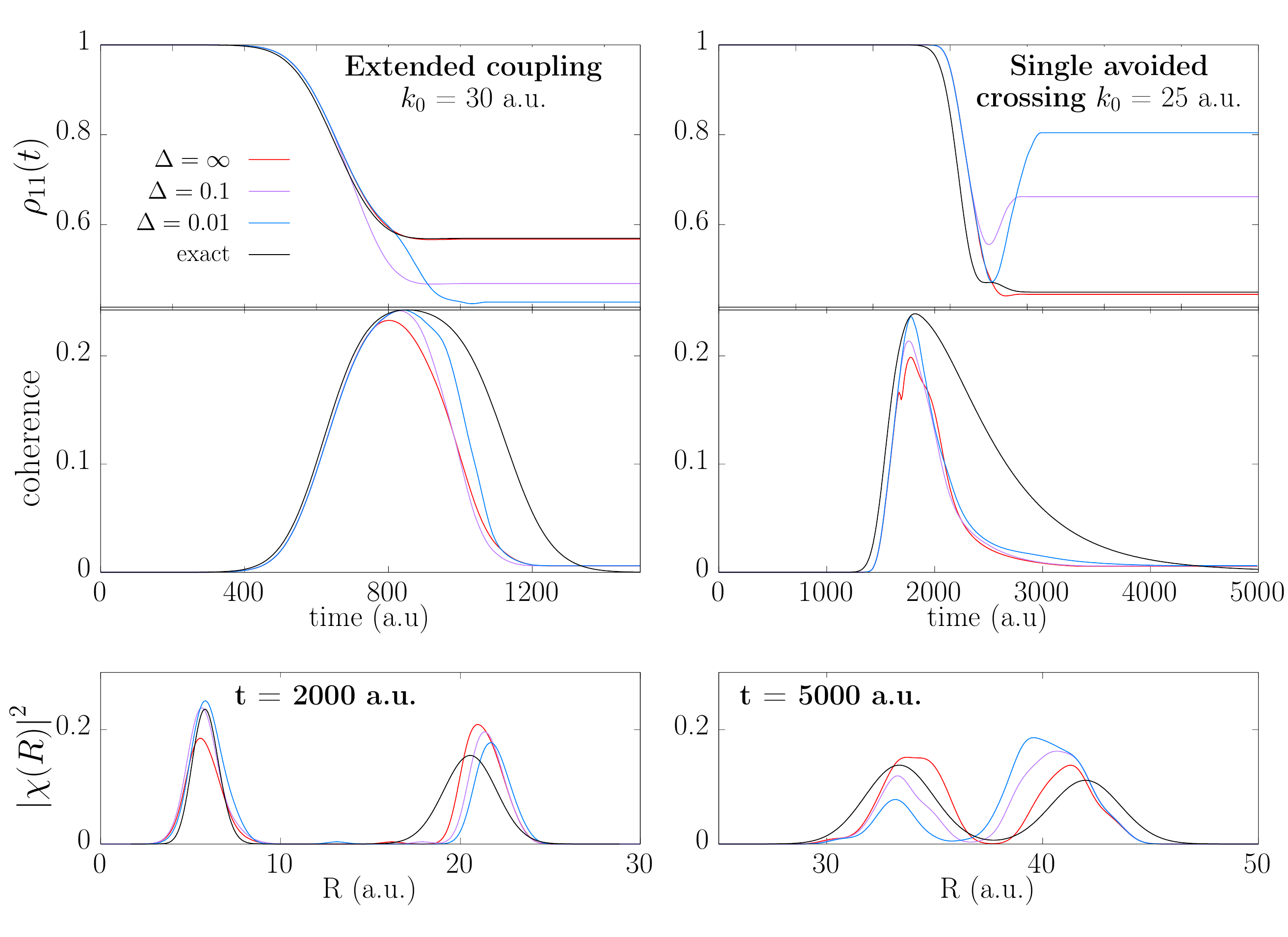}
\caption{Comparisons of the integrated populations, indicators of coherence, and late-time nuclear densities for several values of $\Delta$.}
\label{fig:NAC_cond_plots}
\end{figure}
\end{center}

The results of this approach are plotted in Fig.~\eref{fig:NAC_cond_plots} for the extended-coupling model and for the single avoided crossing model. There we show the population of the initially occupied state as function of time (upper panels), the indicator of coherence (middle panels), and the nuclear density at the final time of the dynamics (lower panels). Although arguably at intermediate times the results are better, overall the results are not as good as when $\mathcal{Q}^{(I)}$ is computed   the original way. The populations are worsened in this approach however the coherences are slightly improved at intermediate times.  This is perhaps not surprising as the purpose of imposing Eq.~\eref{eq:NACcondition} was to ensure correct behavior of the integrated populations. However, the fact that, in the two models considered, the coherences appear to improve at the expense of the integrated populations suggests a rigidity in the definition of the quantum momentum that could be softened with improvements that are currently under investigation. 

\section{A comparison of decoherence times}
\label{sec:decoherencetime}
A well-known problem of both the Ehrenfest and surface-hopping methods was realized soon after they were first introduced: they remained unphysically coherent after passing through an avoided crossing. 
Decoherence corrections have been imposed to attempt to fix the over-coherence problem in the original fewest-switches surface-hopping (FSSH) algorithm~\cite{SBPR96,PR97,SOL13,WAP16,ZNJT04,GP07}, while also alternative surface-hopping methods have been derived whose equations naturally include decoherence~\cite{M16b,GT17}.     

Recently a decoherence correction was derived beginning from the quantum-classical Liouville-equation~\cite{KC99,SOL13} and a simple  approximation of this was shown to be equivalent to the A-FSSH of Refs.~\cite{SS11,LS12}. The correction has the form of a decoherence rate, initially defined via the decay of the off-diagonal electronic density-matrix elements in the absence of NACs,
\begin{equation}
\tau^{-1} = -\frac{d}{dt}\ln|\rho_{12}|
\label{eq:tau-raw}
\end{equation}
Subsequent analysis~\cite{SS11,LS12} then led eventually to a decoherence time defined for each trajectory~\cite{SS11,LS12,SOL13}, via
\ben
\tau_{\rm AFSSH}^{-1} = \frac{1}{2\hbar}(\delta R_1 - \delta R_2)(F_1 - F_2)
\label{eq:tau-AFSSH}
\een
(written for the two-level case), where $F_1-F_2$ is the difference in the BO forces at the current position of the trajectory, and $\delta R_1-\delta R_2$ is the difference in the position-moments of the running trajectory and an auxiliary one that evolves on the non-active surface. 
In practise, this decoherence time is used in a fewest-collapses stochastic way within FSSH that determines whether the electronic coefficient collapses to a pure BO state or not.

The CT-MQC algorithm in the absence of NACs, on the other hand, gives the equation (see Eq.~\eref{eq:offdiag})
\ben
\frac{d}{dt}\rho_{12} =  i(\epsilon_1-\epsilon_2)\rho_{12}
 - \frac{\mathcal{Q}}{M}(\rho_{11} - \rho_{22})(f_1-f_2) \rho_{12}  
 \een
from which we identify an equivalent CT-MQC decoherence rate through Eq.~\eref{eq:tau-raw} as
\ben
\tau^{-1}_{\rm CT-MQC} = (\rho_{11}-\rho_{22}) (f_1-f_2) \mathcal{Q}^{(I)} /M
\label{eq:tau}
\een
which has meaning when $\rho_{12} \neq 0$. 

 We will compare $\tau^{-1}_{\rm AFSSH}$ of Eq.~\eref{eq:tau-AFSSH} and $\tau^{-1}_{\rm CT-MQC}$ of Eq.~\eref{eq:tau} using both the CT-MQC data and exact quantum wavepacket data. 
Although Eq.~\eref{eq:tau} is applied on top of Ehrenfest dynamics, while Eq.~\eref{eq:tau-AFSSH} is a correction to surface-hopping dynamics, it is instructive to compare their structures. A salient point is that $\tau_{\rm CT-MQC}$ depends on the force-difference integrated along the trajectory, while $\tau_{\rm AFSSH}$ depends on the instantaneous force-difference. 

\begin{center}
\begin{figure}
\includegraphics[width=.5\textwidth, height = 0.2\textwidth]{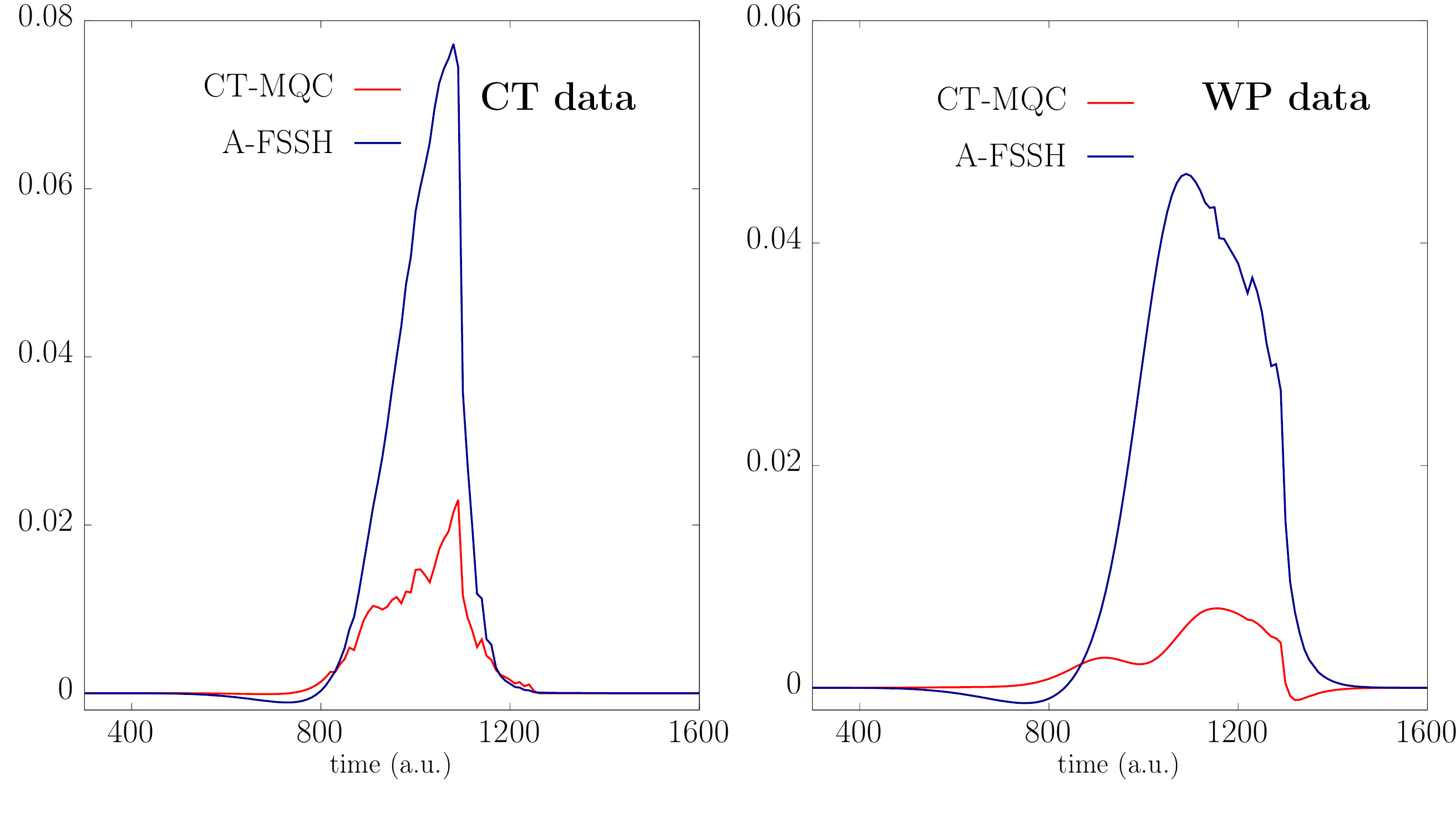}
\caption{Comparison of trajectory-averaged $\tau^{-1}_{\rm CT-MQC}$ and $\tau^{-1}_{\rm AFSSH}$ computed from CT-MQC data  (left) and quantum wavepacket data (right).}
\label{fig:T3k30_tau}
\end{figure}
\end{center}

Next we explain how we calculate the four decoherence rates.
\begin{enumerate}[(i)]
\item $\tau^{-1}_{\rm AFSSH}[{\rm CT-MQC}]$: 
To obtain an effective $\tau_{\rm AFSSH}$ from the  CT-MQC  data we find the effective difference in the position moments, $\delta R_1-\delta R_2$ from the moments of the projected nuclear density distributions in the following way (see also Ref.~\cite{AMAG16}):
\begin{equation}
\langle R_j \rangle = \frac{1}{\mathcal{N}_j} \frac{1}{N}\sum_I \rho^{(I)}_{jj} R^{(I)}\,, \; \mathcal{N}_j =  \frac{1}{N}\sum_I \rho^{(I)}_{jj}
\label{eq:moment-CT-MQC}
\end{equation}
with $j$ labelling the electronic states. Then,
\ben
\tau^{-1}_{\rm AFSSH}[{\rm CT-MQC}] = \frac{1}{2}(F_{1}^{(I)}-F_{2}^{(I)})\left( \langle R_1 \rangle - \langle R_2 \rangle\right)
\een
would give an analog to the A-FSSH decoherence time for trajectory $I$ evaluated on the CT-MQC data, while taking an average of the force difference in the expression above would give a trajectory-averaged decoherence time. 
\item$\tau^{-1}_{\rm AFSSH}[{\rm exact}] $:
To obtain $\tau_{\rm AFSSH}$ from the exact quantum wavepacket data, we define
\ben
\tau^{-1}_{\rm AFSSH}[{\rm exact}] = \frac{1}{2}(-\nabla\epsilon_{1}(R)+\nabla\epsilon_2(R))\left( \langle R_1 \rangle - \langle R_2 \rangle \right)
\een
where the moment difference is this time evaluated from the exact projections of the nuclear wavepackets on the corresponding electronic state, i.e. replacing Eq.~\eref{eq:moment-CT-MQC} with $\langle R_j \rangle =\int |\chi_j(R)\vert^2 R dR/\int |\chi_j(R)|^2 dR$ where $\chi_{j}(R) = \int \Phi^{(j)}_{\rm BO}(r)\Psi(r,R,t)dr$. 
\item$\tau^{-1}_{\rm CT-MQC}$: This is simply obtained directly from Eq.~\eref{eq:tau}, evaluated along the CT-MQC algorithm. 
\item$\tau^{-1}_{\rm CT-MQC}[{\rm exact}] $: Here we replace all quantities in Eq.~\eref{eq:tau} with their values obtained from the exact quantum calculation for the coupled system, i.e. 
\begin{align}
\tau^{-1}_{\rm CT-MQC}[{\rm exact}] = & (\rho_{11} - \rho_{22})  \notag{} \\
&\times (- t \nabla \epsilon^{BO}_{1}(R) + t \nabla\epsilon^{BO}_{2}(R)) \mathcal{Q}(R,t)/M
\end{align}
where $\mathcal{Q}(R,t)= -\nabla_R |\chi(R,t) |/|\chi(R,t)|$ is computed from the exact nuclear wavefunction, and $\rho_{ll} =\vert \int \Phi^{(l)}_{\rm BO,R}(r)\Psi(r,R,t) dr \vert^2/|\chi(R,t)|^2 $. 
\end{enumerate}
In Fig.~\eref{fig:T3k30_tau} we plot these four decoherence rates for the extended-coupling model with $k_0 = 30$~a.u.;  trajectory-averaged (for (i) and (iii)) or wavepacket-averaged (for (ii) and (iv)), are obtained by sum over the trajectories or integration over the nuclear wavepacket, respectively. Since the decoherence rates only have meaning when $\rho_{12}$ is zero, we manually set them to zero when the populations come within a threshold (of 0.005) of $1$ or $0$.

All the four decoherence rates show a similar structure in that they begin to turn on around $t = 800$~a.u. when the system is exiting the NAC region, and its populations beginning to stabilize. The decoherence rate $\tau^{-1}_{\rm CT-MQC}$ rises and  falls back to zero by $t = 1100$~a.u. when evaluated on the CT-MQC data, and  is consistent with the behavior of the coherence at that time (c.f. Fig.~\ref{fig:highk_coh_pop}), while when evaluated on the exact data, persists a little longer, consistent also with the exact coherence. The rate $\tau^{-1}_{\rm AFSSH}$ rises to a much larger value when evaluated on the exact data, and still larger when evaluated on the CT-MQC data, and begins to fall to zero at about the same time in both cases. It should be borne in mind however that in practise $\tau_{\mathrm{AFSSH}}$ is used in a stochastic scheme, where it determines the probability of collapse of a coefficient; if a coefficient is already essentially zero with all the population in one state, such as beyond about $t = 1300$~a.u. (exact case) and $t  = 1100$~a.u. (CT-MQC case), there is no collapse anyway and $\tau_{\rm AFSSH}$ does not actually have any effect. 
Whether the effect of its much larger magnitude is also somehow mitigated in a stochastic scheme is an open question. Further, it is intended to be evaluated on surface-hopping dynamics, so one must be wary of making too quantitative a comparison, but if surface-hopping dynamics mimics the exact dynamics well enough, then this should not be a major factor in the difference.

\section{Conclusions and Outlook}
\label{sec:conclusions}
The purpose of this work was to explore the approximations introduced in deriving CT-MQC, in order to highlight the importance of each term of 
the CT-MQC equations, and in particular shed light on how the coupled-trajectory terms actually operate in practise. In previous work~\cite{MAG15,AMAG16}, only the overall effect of the CT-MQC algorithm has been illustrated by computation of observables and comparisons with other quantum-classical methods. Here, we deepened the analysis of the algorithm by thoroughly investigating the quantities that appear in the CT-MQC equations.

We found that the term coupling the trajectories in the electronic equation is of paramount importance in obtaining accurate dynamics, and discussed how and why it causes branching of populations and ensuing wavepacket splitting. The term that couples the trajectories in the nuclear equation is less critical, and in many cases, reasonably accurate dynamics could be obtained by neglecting it. We demonstrated our analysis on fully time-resolved and spatially-resolved nuclear densities and electronic state coefficients, as well as the more conventionally used branching ratios. The latter only give final-time information and limited spatial resolution, as well as the time-resolved trajectory-averaged populations and coherences which do not indicate spatial information. We could extract an effective decoherence time from the algorithm and compare with that of A-FSSH. 

Our analysis additionally has the potential to shed light into the possibility of proposing more efficient or more accurate approximations.  The realization that decoherence and branching are closely reproduced by including the coupled-trajectory term only in the electronic equation, opens the possibility of developing simpler mixed quantum-classical methods in the future. We discussed some possibilities of alternative computation of the quantum momentum, and more refinements of this quantity are on-going.

%


\begin{acknowledgments}
Financial support from the US National Science Foundation
CHE-1566197 (G.G.) and the Department of Energy, Office
of Basic Energy Sciences, Division of Chemical Sciences,
Geosciences and Biosciences under Award DE-SC0015344 (N.T.M.), and a Research Corporation For Science Advancement Cottrell Scholar Seed Award,  are gratefully acknowledged. 
\end{acknowledgments}

\appendix

\section{Tully Models}
The Hamiltonian defining the model systems studied here is, in the diabatic basis, of the form
\begin{align}
\hat H = \frac{\hat P}{2M} + \left(
\begin{array}{cc}
H_{11}(R) & H_{12}(R) \\ 
H_{12}(R) & H_{22}(R)
\end{array}
\right).
\end{align}
The nuclear mass is set to 2000~a.u.; the elements of the $2\times2$ electronic Hamiltonian depend on nuclear positions $R$. Their expressions yield different non-adiabatic coupling situations, as described below.

\subsection{Extended coupling}
\begin{align}
 H_{11}(R)& = A\notag{} \\
H_{12}(R) &= \begin{cases} 
   B e^{C R} & {R \leq 0} \\
     B \left(2-e^{-C R}\right) & R > 0
   \end{cases}\\
H_{21}(R) &= H_{12}(R)\\
H_{22}(R) &=-H_{11}(R)
\end{align}
with $A=0.0006$, $B=0.1$, and $C=0.9$.

\subsection{Single avoided crossing}
\begin{align}
H_{11}(R) & = A \mathrm{tanh}(B R) \notag{} \\
H_{12}(R) & = C e^{-D R^2} \notag{} \\
H_{21}(R) & = H_{12}(R)\notag{} \\
H_{22}(R) & = -H_{11}(R)
\end{align}
with $A=0.03$, $B=0.4$, $C=0.005$, $D=0.3$.

\subsection{Double arch}
\begin{align}
 H_{11}(R)& = a \notag{} \\
H_{12}(R) &= \begin{cases} 
	 B \left(-e^{C(R-D)}+e^{C(R+D)}\right), & R\leq -D \notag{} \\
          B \left( e^{-C(R-D)}- e^{-C(R+D)} \right)& R \geq D \notag{} \\
           B \left( 2 - e^{C(R-D)}-e^{-C(R+D)} \right) & -D<R<D \notag{}
   \end{cases}\\
H_{21}(R) &= H_{12}(R) \notag{} \\
H_{22}(R) &=-H_{11}(R)
\end{align}
with $A=0.0006$, $B=0.1$, $C=0.9$, $D=4.0$.

\subsection{Dual avoided crossing}
\begin{align}
H_{11}(R) & = 0 \notag{} \\
H_{12}(R) & = C e^{-D R^2}\notag{} \\
H_{21}(R) & = H_{12}(R)\notag{} \\
H_{22}(R) & = -A e^{-B R^2}+E
\end{align}
with $A=0.1$, $B=0.28$, $C=0.015$, $D=0.06$, $E=0.05$.

\section{Computational Details}
In the quantum-classical simulations, the equation of motion for the BO coefficients, Eq.~\eref{eq:CT-MQCeqnse}, is propagated using fourth-order Runge-Kutta algorithm, with the nuclear degrees of freedom, Eq.~\eref{eq:CT-MQCeqnsn}, evolved using the velocity-Verlet algorithm. A time step of $dt=0.1$~a.u.  is used for propagating both the electronic and nuclear equations.

In CT-MQC and CTe-MQC,  decoherence is deemed to have occured when a trajectory is completely on one surface or another. However, computationally the electronic matrix elements $\rho_{ll}^{(I)}$ will never reach exactly $1$ or $0$ after the interaction, and so a cutoff is imposed: when $\rho_{ll}^{(I)}(t) > 0.995$ or $\rho_{ll}^{(I)}(t)<0.005$ the integrated force difference is set to zero. This ensures the correct removal of any history dependence accumulated in the integrated force, as expected for a completely decohered wave packet.

Exact wavepacket propagation is performed by evolving the initial state with a time step of $dt=0.1$~a.u. in the diabatic basis using the split-operator technique~\cite{spo}.

\bibliography{./ref_na}

\begin{thebibliography}{74}%
\makeatletter
\providecommand \@ifxundefined [1]{%
 \@ifx{#1\undefined}
}%
\providecommand \@ifnum [1]{%
 \ifnum #1\expandafter \@firstoftwo
 \else \expandafter \@secondoftwo
 \fi
}%
\providecommand \@ifx [1]{%
 \ifx #1\expandafter \@firstoftwo
 \else \expandafter \@secondoftwo
 \fi
}%
\providecommand \natexlab [1]{#1}%
\providecommand \enquote  [1]{``#1''}%
\providecommand \bibnamefont  [1]{#1}%
\providecommand \bibfnamefont [1]{#1}%
\providecommand \citenamefont [1]{#1}%
\providecommand \href@noop [0]{\@secondoftwo}%
\providecommand \href [0]{\begingroup \@sanitize@url \@href}%
\providecommand \@href[1]{\@@startlink{#1}\@@href}%
\providecommand \@@href[1]{\endgroup#1\@@endlink}%
\providecommand \@sanitize@url [0]{\catcode `\\12\catcode `\$12\catcode
  `\&12\catcode `\#12\catcode `\^12\catcode `\_12\catcode `\%12\relax}%
\providecommand \@@startlink[1]{}%
\providecommand \@@endlink[0]{}%
\providecommand \url  [0]{\begingroup\@sanitize@url \@url }%
\providecommand \@url [1]{\endgroup\@href {#1}{\urlprefix }}%
\providecommand \urlprefix  [0]{URL }%
\providecommand \Eprint [0]{\href }%
\providecommand \doibase [0]{http://dx.doi.org/}%
\providecommand \selectlanguage [0]{\@gobble}%
\providecommand \bibinfo  [0]{\@secondoftwo}%
\providecommand \bibfield  [0]{\@secondoftwo}%
\providecommand \translation [1]{[#1]}%
\providecommand \BibitemOpen [0]{}%
\providecommand \bibitemStop [0]{}%
\providecommand \bibitemNoStop [0]{.\EOS\space}%
\providecommand \EOS [0]{\spacefactor3000\relax}%
\providecommand \BibitemShut  [1]{\csname bibitem#1\endcsname}%
\let\auto@bib@innerbib\@empty
\bibitem [{\citenamefont {Prezhdo}\ \emph {et~al.}(2009)\citenamefont
  {Prezhdo}, \citenamefont {Duncan},\ and\ \citenamefont {Prezhdo}}]{PDP09}%
  \BibitemOpen
  \bibfield  {author} {\bibinfo {author} {\bibfnamefont {O.~V.}\ \bibnamefont
  {Prezhdo}}, \bibinfo {author} {\bibfnamefont {W.~R.}\ \bibnamefont {Duncan}},
  \ and\ \bibinfo {author} {\bibfnamefont {V.~V.}\ \bibnamefont {Prezhdo}},\
  }\href@noop {} {\bibfield  {journal} {\bibinfo  {journal} {Progress in
  Surface Science}\ }\textbf {\bibinfo {volume} {84}},\ \bibinfo {pages} {30 }
  (\bibinfo {year} {2009})}\BibitemShut {NoStop}%
\bibitem [{\citenamefont {van Leeuwen}\ \emph {et~al.}(2017)\citenamefont {van
  Leeuwen}, \citenamefont {Lubbe}, \citenamefont {Stacko}, \citenamefont
  {Wezenberg},\ and\ \citenamefont {Feringa}}]{Feringa_2017}%
  \BibitemOpen
  \bibfield  {author} {\bibinfo {author} {\bibfnamefont {T.}~\bibnamefont {van
  Leeuwen}}, \bibinfo {author} {\bibfnamefont {A.~S.}\ \bibnamefont {Lubbe}},
  \bibinfo {author} {\bibfnamefont {P.}~\bibnamefont {Stacko}}, \bibinfo
  {author} {\bibfnamefont {S.~J.}\ \bibnamefont {Wezenberg}}, \ and\ \bibinfo
  {author} {\bibfnamefont {B.~L.}\ \bibnamefont {Feringa}},\ }\href@noop {}
  {\bibfield  {journal} {\bibinfo  {journal} {Nature Reviews Chemistry}\
  }\textbf {\bibinfo {volume} {1}},\ \bibinfo {pages} {0096} (\bibinfo {year}
  {2017})}\BibitemShut {NoStop}%
\bibitem [{\citenamefont {Levine}\ and\ \citenamefont
  {Martínez}(2007)}]{LM07}%
  \BibitemOpen
  \bibfield  {author} {\bibinfo {author} {\bibfnamefont {B.~G.}\ \bibnamefont
  {Levine}}\ and\ \bibinfo {author} {\bibfnamefont {T.~J.}\ \bibnamefont
  {Martínez}},\ }\href {\doibase 10.1146/annurev.physchem.57.032905.104612}
  {\bibfield  {journal} {\bibinfo  {journal} {Annual Review of Physical
  Chemistry}\ }\textbf {\bibinfo {volume} {58}},\ \bibinfo {pages} {613}
  (\bibinfo {year} {2007})},\ \bibinfo {note} {pMID: 17291184},\ \Eprint
  {http://arxiv.org/abs/https://doi.org/10.1146/annurev.physchem.57.032905.104612}
  {https://doi.org/10.1146/annurev.physchem.57.032905.104612} \BibitemShut
  {NoStop}%
\bibitem [{\citenamefont {van~der Horst}\ \emph {et~al.}(2004)\citenamefont
  {van~der Horst}, \citenamefont {Hellingwerf},\ and\ \citenamefont
  {J.}}]{Horst2004}%
  \BibitemOpen
  \bibfield  {author} {\bibinfo {author} {\bibfnamefont {M.~A.}\ \bibnamefont
  {van~der Horst}}, \bibinfo {author} {\bibnamefont {Hellingwerf}}, \ and\
  \bibinfo {author} {\bibfnamefont {K.}~\bibnamefont {J.}},\ }\href {\doibase
  10.1021/ar020219d} {\bibfield  {journal} {\bibinfo  {journal} {Accounts of
  Chemical Research}\ }\textbf {\bibinfo {volume} {37}},\ \bibinfo {pages} {13}
  (\bibinfo {year} {2004})},\ \bibinfo {note} {pMID: 14730990},\ \Eprint
  {http://arxiv.org/abs/https://doi.org/10.1021/ar020219d}
  {https://doi.org/10.1021/ar020219d} \BibitemShut {NoStop}%
\bibitem [{\citenamefont {L{\'e}pine}\ \emph {et~al.}(2014)\citenamefont
  {L{\'e}pine}, \citenamefont {Ivanov},\ and\ \citenamefont
  {Vrakking}}]{Vrakking_NP2014}%
  \BibitemOpen
  \bibfield  {author} {\bibinfo {author} {\bibfnamefont {F.}~\bibnamefont
  {L{\'e}pine}}, \bibinfo {author} {\bibfnamefont {M.~Y.}\ \bibnamefont
  {Ivanov}}, \ and\ \bibinfo {author} {\bibfnamefont {M.~J.~J.}\ \bibnamefont
  {Vrakking}},\ }\href@noop {} {\bibfield  {journal} {\bibinfo  {journal}
  {Nature Photonics}\ }\textbf {\bibinfo {volume} {8}},\ \bibinfo {pages} {195}
  (\bibinfo {year} {2014})}\BibitemShut {NoStop}%
\bibitem [{\citenamefont {Nisoli}\ \emph {et~al.}(2017)\citenamefont {Nisoli},
  \citenamefont {Decleva}, \citenamefont {Calegari}, \citenamefont {Palacios},\
  and\ \citenamefont {Martín}}]{Martin2017}%
  \BibitemOpen
  \bibfield  {author} {\bibinfo {author} {\bibfnamefont {M.}~\bibnamefont
  {Nisoli}}, \bibinfo {author} {\bibfnamefont {P.}~\bibnamefont {Decleva}},
  \bibinfo {author} {\bibfnamefont {F.}~\bibnamefont {Calegari}}, \bibinfo
  {author} {\bibfnamefont {A.}~\bibnamefont {Palacios}}, \ and\ \bibinfo
  {author} {\bibfnamefont {F.}~\bibnamefont {Martín}},\ }\href {\doibase
  10.1021/acs.chemrev.6b00453} {\bibfield  {journal} {\bibinfo  {journal}
  {Chemical Reviews}\ }\textbf {\bibinfo {volume} {117}},\ \bibinfo {pages}
  {10760} (\bibinfo {year} {2017})},\ \bibinfo {note} {pMID: 28488433},\
  \Eprint {http://arxiv.org/abs/https://doi.org/10.1021/acs.chemrev.6b00453}
  {https://doi.org/10.1021/acs.chemrev.6b00453} \BibitemShut {NoStop}%
\bibitem [{\citenamefont {{McLachlan}}(1964)}]{M64}%
  \BibitemOpen
  \bibfield  {author} {\bibinfo {author} {\bibfnamefont {A.~D.}\ \bibnamefont
  {{McLachlan}}},\ }\href {\doibase 10.1080/00268976400100041} {\bibfield
  {journal} {\bibinfo  {journal} {Molecular Physics}\ }\textbf {\bibinfo
  {volume} {8}},\ \bibinfo {pages} {39} (\bibinfo {year} {1964})}\BibitemShut
  {NoStop}%
\bibitem [{\citenamefont {Tully}(1998)}]{T98}%
  \BibitemOpen
  \bibfield  {author} {\bibinfo {author} {\bibfnamefont {J.~C.}\ \bibnamefont
  {Tully}},\ }\href@noop {} {\bibfield  {journal} {\bibinfo  {journal} {Faraday
  Discuss.}\ }\textbf {\bibinfo {volume} {110}},\ \bibinfo {pages} {407}
  (\bibinfo {year} {1998})}\BibitemShut {NoStop}%
\bibitem [{\citenamefont {Tully}\ and\ \citenamefont {Preston}(1971)}]{TP71}%
  \BibitemOpen
  \bibfield  {author} {\bibinfo {author} {\bibfnamefont {J.~C.}\ \bibnamefont
  {Tully}}\ and\ \bibinfo {author} {\bibfnamefont {R.}~\bibnamefont
  {Preston}},\ }\href@noop {} {\bibfield  {journal} {\bibinfo  {journal} {J.
  Chem. Phys.}\ }\textbf {\bibinfo {volume} {55}},\ \bibinfo {pages} {562}
  (\bibinfo {year} {1971})}\BibitemShut {NoStop}%
\bibitem [{\citenamefont {Tully}(1990)}]{T90}%
  \BibitemOpen
  \bibfield  {author} {\bibinfo {author} {\bibfnamefont {J.~C.}\ \bibnamefont
  {Tully}},\ }\href@noop {} {\bibfield  {journal} {\bibinfo  {journal} {J.
  Chem. Phys.}\ }\textbf {\bibinfo {volume} {93}},\ \bibinfo {pages} {1061}
  (\bibinfo {year} {1990})}\BibitemShut {NoStop}%
\bibitem [{\citenamefont {Subotnik}\ \emph {et~al.}(2016)\citenamefont
  {Subotnik}, \citenamefont {Jain}, \citenamefont {Landry}, \citenamefont
  {Petit}, \citenamefont {Ouyang},\ and\ \citenamefont
  {Bellonzi}}]{Subotnik_ARPC2016}%
  \BibitemOpen
  \bibfield  {author} {\bibinfo {author} {\bibfnamefont {J.~E.}\ \bibnamefont
  {Subotnik}}, \bibinfo {author} {\bibfnamefont {A.}~\bibnamefont {Jain}},
  \bibinfo {author} {\bibfnamefont {B.}~\bibnamefont {Landry}}, \bibinfo
  {author} {\bibfnamefont {A.}~\bibnamefont {Petit}}, \bibinfo {author}
  {\bibfnamefont {W.}~\bibnamefont {Ouyang}}, \ and\ \bibinfo {author}
  {\bibfnamefont {N.}~\bibnamefont {Bellonzi}},\ }\href@noop {} {\bibfield
  {journal} {\bibinfo  {journal} {Ann. Rev. Phys. Chem.}\ }\textbf {\bibinfo
  {volume} {67}},\ \bibinfo {pages} {387} (\bibinfo {year} {2016})}\BibitemShut
  {NoStop}%
\bibitem [{\citenamefont {Wang}\ \emph
  {et~al.}(2016{\natexlab{a}})\citenamefont {Wang}, \citenamefont {Akimov},\
  and\ \citenamefont {Prezhdo}}]{Prezhdo_JPCL2016}%
  \BibitemOpen
  \bibfield  {author} {\bibinfo {author} {\bibfnamefont {L.}~\bibnamefont
  {Wang}}, \bibinfo {author} {\bibfnamefont {A.}~\bibnamefont {Akimov}}, \ and\
  \bibinfo {author} {\bibfnamefont {O.~V.}\ \bibnamefont {Prezhdo}},\
  }\href@noop {} {\bibfield  {journal} {\bibinfo  {journal} {J. Phys. Chem.
  Lett.}\ }\textbf {\bibinfo {volume} {7}},\ \bibinfo {pages} {2100} (\bibinfo
  {year} {2016}{\natexlab{a}})}\BibitemShut {NoStop}%
\bibitem [{\citenamefont {Lu}\ and\ \citenamefont {Zhou}(2017)}]{LZ17}%
  \BibitemOpen
  \bibfield  {author} {\bibinfo {author} {\bibfnamefont {J.}~\bibnamefont
  {Lu}}\ and\ \bibinfo {author} {\bibfnamefont {Z.}~\bibnamefont {Zhou}},\
  }\href@noop {} {\bibfield  {journal} {\bibinfo  {journal} {Math. Comp.}\ }
  (\bibinfo {year} {2017})}\BibitemShut {NoStop}%
\bibitem [{\citenamefont {Min}\ \emph {et~al.}(2015)\citenamefont {Min},
  \citenamefont {Agostini},\ and\ \citenamefont {Gross}}]{MAG15}%
  \BibitemOpen
  \bibfield  {author} {\bibinfo {author} {\bibfnamefont {S.~K.}\ \bibnamefont
  {Min}}, \bibinfo {author} {\bibfnamefont {F.}~\bibnamefont {Agostini}}, \
  and\ \bibinfo {author} {\bibfnamefont {E.~K.~U.}\ \bibnamefont {Gross}},\
  }\href@noop {} {\bibfield  {journal} {\bibinfo  {journal} {Phys. Rev. Lett.}\
  }\textbf {\bibinfo {volume} {115}},\ \bibinfo {pages} {073001} (\bibinfo
  {year} {2015})}\BibitemShut {NoStop}%
\bibitem [{\citenamefont {Abedi}\ \emph {et~al.}(2010)\citenamefont {Abedi},
  \citenamefont {Maitra},\ and\ \citenamefont {Gross}}]{AMG10}%
  \BibitemOpen
  \bibfield  {author} {\bibinfo {author} {\bibfnamefont {A.}~\bibnamefont
  {Abedi}}, \bibinfo {author} {\bibfnamefont {N.~T.}\ \bibnamefont {Maitra}}, \
  and\ \bibinfo {author} {\bibfnamefont {E.~K.~U.}\ \bibnamefont {Gross}},\
  }\href@noop {} {\bibfield  {journal} {\bibinfo  {journal} {Phys. Rev. Lett.}\
  }\textbf {\bibinfo {volume} {105}},\ \bibinfo {pages} {123002} (\bibinfo
  {year} {2010})}\BibitemShut {NoStop}%
\bibitem [{\citenamefont {Abedi}\ \emph {et~al.}(2012)\citenamefont {Abedi},
  \citenamefont {Maitra},\ and\ \citenamefont {Gross}}]{AMG12}%
  \BibitemOpen
  \bibfield  {author} {\bibinfo {author} {\bibfnamefont {A.}~\bibnamefont
  {Abedi}}, \bibinfo {author} {\bibfnamefont {N.~T.}\ \bibnamefont {Maitra}}, \
  and\ \bibinfo {author} {\bibfnamefont {E.~K.~U.}\ \bibnamefont {Gross}},\
  }\href@noop {} {\bibfield  {journal} {\bibinfo  {journal} {J. Chem. Phys.}\
  }\textbf {\bibinfo {volume} {137}},\ \bibinfo {pages} {22A530} (\bibinfo
  {year} {2012})}\BibitemShut {NoStop}%
\bibitem [{\citenamefont {Agostini}\ \emph {et~al.}(2016)\citenamefont
  {Agostini}, \citenamefont {Min}, \citenamefont {Abedi},\ and\ \citenamefont
  {Gross}}]{AMAG16}%
  \BibitemOpen
  \bibfield  {author} {\bibinfo {author} {\bibfnamefont {F.}~\bibnamefont
  {Agostini}}, \bibinfo {author} {\bibfnamefont {S.~K.}\ \bibnamefont {Min}},
  \bibinfo {author} {\bibfnamefont {A.}~\bibnamefont {Abedi}}, \ and\ \bibinfo
  {author} {\bibfnamefont {E.~K.~U.}\ \bibnamefont {Gross}},\ }\href {\doibase
  10.1021/acs.jctc.5b01180} {\bibfield  {journal} {\bibinfo  {journal} {Journal
  of Chemical Theory and Computation}\ }\textbf {\bibinfo {volume} {12}},\
  \bibinfo {pages} {2127} (\bibinfo {year} {2016})},\ \bibinfo {note} {pMID:
  27030209},\ \Eprint
  {http://arxiv.org/abs/http://dx.doi.org/10.1021/acs.jctc.5b01180}
  {http://dx.doi.org/10.1021/acs.jctc.5b01180} \BibitemShut {NoStop}%
\bibitem [{\citenamefont {Min}\ \emph {et~al.}(2017)\citenamefont {Min},
  \citenamefont {Agostini}, \citenamefont {Tavernelli},\ and\ \citenamefont
  {Gross}}]{MATG17}%
  \BibitemOpen
  \bibfield  {author} {\bibinfo {author} {\bibfnamefont {S.~K.}\ \bibnamefont
  {Min}}, \bibinfo {author} {\bibfnamefont {F.}~\bibnamefont {Agostini}},
  \bibinfo {author} {\bibfnamefont {I.}~\bibnamefont {Tavernelli}}, \ and\
  \bibinfo {author} {\bibfnamefont {E.~K.~U.}\ \bibnamefont {Gross}},\ }\href
  {\doibase 10.1021/acs.jpclett.7b01249} {\bibfield  {journal} {\bibinfo
  {journal} {The Journal of Physical Chemistry Letters}\ }\textbf {\bibinfo
  {volume} {8}},\ \bibinfo {pages} {3048} (\bibinfo {year} {2017})},\ \bibinfo
  {note} {pMID: 28618782},\ \Eprint
  {http://arxiv.org/abs/http://dx.doi.org/10.1021/acs.jpclett.7b01249}
  {http://dx.doi.org/10.1021/acs.jpclett.7b01249} \BibitemShut {NoStop}%
\bibitem [{\citenamefont {Subotnik}\ and\ \citenamefont {Shenvi}(2011)}]{SS11}%
  \BibitemOpen
  \bibfield  {author} {\bibinfo {author} {\bibfnamefont {J.~E.}\ \bibnamefont
  {Subotnik}}\ and\ \bibinfo {author} {\bibfnamefont {N.}~\bibnamefont
  {Shenvi}},\ }\href@noop {} {\bibfield  {journal} {\bibinfo  {journal} {J.
  Chem. Phys.}\ }\textbf {\bibinfo {volume} {134}},\ \bibinfo {pages} {244114}
  (\bibinfo {year} {2011})}\BibitemShut {NoStop}%
\bibitem [{\citenamefont {Landry}\ and\ \citenamefont {Subotnik}(2012)}]{LS12}%
  \BibitemOpen
  \bibfield  {author} {\bibinfo {author} {\bibfnamefont {B.~R.}\ \bibnamefont
  {Landry}}\ and\ \bibinfo {author} {\bibfnamefont {J.~E.}\ \bibnamefont
  {Subotnik}},\ }\href {\doibase 10.1063/1.4733675} {\bibfield  {journal}
  {\bibinfo  {journal} {The Journal of Chemical Physics}\ }\textbf {\bibinfo
  {volume} {137}},\ \bibinfo {pages} {22A513} (\bibinfo {year} {2012})},\
  \Eprint {http://arxiv.org/abs/https://doi.org/10.1063/1.4733675}
  {https://doi.org/10.1063/1.4733675} \BibitemShut {NoStop}%
\bibitem [{\citenamefont {Subotnik}\ \emph {et~al.}(2013)\citenamefont
  {Subotnik}, \citenamefont {Ouyang},\ and\ \citenamefont {Landry}}]{SOL13}%
  \BibitemOpen
  \bibfield  {author} {\bibinfo {author} {\bibfnamefont {J.~E.}\ \bibnamefont
  {Subotnik}}, \bibinfo {author} {\bibfnamefont {W.}~\bibnamefont {Ouyang}}, \
  and\ \bibinfo {author} {\bibfnamefont {B.~R.}\ \bibnamefont {Landry}},\
  }\href@noop {} {\bibfield  {journal} {\bibinfo  {journal} {J. Chem. Phys.}\
  }\textbf {\bibinfo {volume} {139}},\ \bibinfo {pages} {214107} (\bibinfo
  {year} {2013})}\BibitemShut {NoStop}%
\bibitem [{\citenamefont {Alonso}\ \emph {et~al.}(2013)\citenamefont {Alonso},
  \citenamefont {Clemente-Gallardo}, \citenamefont {Echenique-Robba},\ and\
  \citenamefont {Jover-Galtier}}]{ACEJ13}%
  \BibitemOpen
  \bibfield  {author} {\bibinfo {author} {\bibfnamefont {J.~L.}\ \bibnamefont
  {Alonso}}, \bibinfo {author} {\bibfnamefont {J.}~\bibnamefont
  {Clemente-Gallardo}}, \bibinfo {author} {\bibfnamefont {P.}~\bibnamefont
  {Echenique-Robba}}, \ and\ \bibinfo {author} {\bibfnamefont {J.~A.}\
  \bibnamefont {Jover-Galtier}},\ }\href {\doibase 10.1063/1.4818521}
  {\bibfield  {journal} {\bibinfo  {journal} {The Journal of Chemical Physics}\
  }\textbf {\bibinfo {volume} {139}},\ \bibinfo {pages} {087101} (\bibinfo
  {year} {2013})},\ \Eprint
  {http://arxiv.org/abs/https://doi.org/10.1063/1.4818521}
  {https://doi.org/10.1063/1.4818521} \BibitemShut {NoStop}%
\bibitem [{\citenamefont {Abedi}\ \emph
  {et~al.}(2013{\natexlab{a}})\citenamefont {Abedi}, \citenamefont {Maitra},\
  and\ \citenamefont {Gross}}]{AMG13}%
  \BibitemOpen
  \bibfield  {author} {\bibinfo {author} {\bibfnamefont {A.}~\bibnamefont
  {Abedi}}, \bibinfo {author} {\bibfnamefont {N.~T.}\ \bibnamefont {Maitra}}, \
  and\ \bibinfo {author} {\bibfnamefont {E.~K.~U.}\ \bibnamefont {Gross}},\
  }\href {\doibase 10.1063/1.4818523} {\bibfield  {journal} {\bibinfo
  {journal} {The Journal of Chemical Physics}\ }\textbf {\bibinfo {volume}
  {139}},\ \bibinfo {pages} {087102} (\bibinfo {year} {2013}{\natexlab{a}})},\
  \Eprint {http://arxiv.org/abs/https://doi.org/10.1063/1.4818523}
  {https://doi.org/10.1063/1.4818523} \BibitemShut {NoStop}%
\bibitem [{\citenamefont {Hunter}(1975)}]{H75}%
  \BibitemOpen
  \bibfield  {author} {\bibinfo {author} {\bibfnamefont {G.}~\bibnamefont
  {Hunter}},\ }\href@noop {} {\bibfield  {journal} {\bibinfo  {journal} {Int.
  J. Quantum Chem.}\ }\textbf {\bibinfo {volume} {9}},\ \bibinfo {pages} {237}
  (\bibinfo {year} {1975})}\BibitemShut {NoStop}%
\bibitem [{\citenamefont {Hunter}(1981)}]{H81}%
  \BibitemOpen
  \bibfield  {author} {\bibinfo {author} {\bibfnamefont {G.}~\bibnamefont
  {Hunter}},\ }\href@noop {} {\bibfield  {journal} {\bibinfo  {journal} {Int.
  J. Quantum Chem.}\ }\textbf {\bibinfo {volume} {19}},\ \bibinfo {pages} {755}
  (\bibinfo {year} {1981})}\BibitemShut {NoStop}%
\bibitem [{\citenamefont {Gidopoulos}\ and\ \citenamefont
  {Gross}(2014)}]{GG14}%
  \BibitemOpen
  \bibfield  {author} {\bibinfo {author} {\bibfnamefont {N.~I.}\ \bibnamefont
  {Gidopoulos}}\ and\ \bibinfo {author} {\bibfnamefont {E.~K.~U.}\ \bibnamefont
  {Gross}},\ }\href {\doibase 10.1098/rsta.2013.0059} {\bibfield  {journal}
  {\bibinfo  {journal} {Philosophical Transactions of the Royal Society of
  London A: Mathematical, Physical and Engineering Sciences}\ }\textbf
  {\bibinfo {volume} {372}} (\bibinfo {year} {2014}),\
  10.1098/rsta.2013.0059}\BibitemShut {NoStop}%
\bibitem [{\citenamefont {Suzuki}\ \emph {et~al.}(2014)\citenamefont {Suzuki},
  \citenamefont {Abedi}, \citenamefont {Maitra}, \citenamefont {Yamashita},\
  and\ \citenamefont {Gross}}]{SAMYG14}%
  \BibitemOpen
  \bibfield  {author} {\bibinfo {author} {\bibfnamefont {Y.}~\bibnamefont
  {Suzuki}}, \bibinfo {author} {\bibfnamefont {A.}~\bibnamefont {Abedi}},
  \bibinfo {author} {\bibfnamefont {N.~T.}\ \bibnamefont {Maitra}}, \bibinfo
  {author} {\bibfnamefont {K.}~\bibnamefont {Yamashita}}, \ and\ \bibinfo
  {author} {\bibfnamefont {E.~K.~U.}\ \bibnamefont {Gross}},\ }\href@noop {}
  {\bibfield  {journal} {\bibinfo  {journal} {Phys. Rev. A}\ }\textbf {\bibinfo
  {volume} {89}},\ \bibinfo {pages} {040501(R)} (\bibinfo {year}
  {2014})}\BibitemShut {NoStop}%
\bibitem [{\citenamefont {Khosravi}\ \emph {et~al.}(2015)\citenamefont
  {Khosravi}, \citenamefont {Abedi},\ and\ \citenamefont {Maitra}}]{KAM15}%
  \BibitemOpen
  \bibfield  {author} {\bibinfo {author} {\bibfnamefont {E.}~\bibnamefont
  {Khosravi}}, \bibinfo {author} {\bibfnamefont {A.}~\bibnamefont {Abedi}}, \
  and\ \bibinfo {author} {\bibfnamefont {N.~T.}\ \bibnamefont {Maitra}},\
  }\href@noop {} {\bibfield  {journal} {\bibinfo  {journal} {Phys. Rev. Lett.}\
  }\textbf {\bibinfo {volume} {115}},\ \bibinfo {pages} {263002} (\bibinfo
  {year} {2015})}\BibitemShut {NoStop}%
\bibitem [{\citenamefont {Khosravi}\ \emph {et~al.}(2017)\citenamefont
  {Khosravi}, \citenamefont {Abedi}, \citenamefont {Rubio},\ and\ \citenamefont
  {Maitra}}]{KARM17}%
  \BibitemOpen
  \bibfield  {author} {\bibinfo {author} {\bibfnamefont {E.}~\bibnamefont
  {Khosravi}}, \bibinfo {author} {\bibfnamefont {A.}~\bibnamefont {Abedi}},
  \bibinfo {author} {\bibfnamefont {A.}~\bibnamefont {Rubio}}, \ and\ \bibinfo
  {author} {\bibfnamefont {N.~T.}\ \bibnamefont {Maitra}},\ }\href@noop {}
  {\bibfield  {journal} {\bibinfo  {journal} {Phys. Chem. Chem. Phys.}\
  }\textbf {\bibinfo {volume} {DOI: 10.1039/c6cp08539c}} (\bibinfo {year}
  {2017})}\BibitemShut {NoStop}%
\bibitem [{\citenamefont {Jecko}\ \emph {et~al.}(2015)\citenamefont {Jecko},
  \citenamefont {Sutcliffe},\ and\ \citenamefont {Woolley}}]{JSW15}%
  \BibitemOpen
  \bibfield  {author} {\bibinfo {author} {\bibfnamefont {T.}~\bibnamefont
  {Jecko}}, \bibinfo {author} {\bibfnamefont {B.~T.}\ \bibnamefont
  {Sutcliffe}}, \ and\ \bibinfo {author} {\bibfnamefont {R.~G.}\ \bibnamefont
  {Woolley}},\ }\href@noop {} {\bibfield  {journal} {\bibinfo  {journal} {J.
  Phys. A: Math. Theor.}\ }\textbf {\bibinfo {volume} {48}},\ \bibinfo {pages}
  {445201} (\bibinfo {year} {2015})}\BibitemShut {NoStop}%
\bibitem [{\citenamefont {Meek}\ and\ \citenamefont {Levine}(2016)}]{ML16}%
  \BibitemOpen
  \bibfield  {author} {\bibinfo {author} {\bibfnamefont {G.~A.}\ \bibnamefont
  {Meek}}\ and\ \bibinfo {author} {\bibfnamefont {B.~G.}\ \bibnamefont
  {Levine}},\ }\href {\doibase 10.1063/1.4948786} {\bibfield  {journal}
  {\bibinfo  {journal} {J. Chem. Phys.}\ }\textbf {\bibinfo {volume} {144}},\
  \bibinfo {pages} {184109} (\bibinfo {year} {2016})},\ \Eprint
  {http://arxiv.org/abs/https://doi.org/10.1063/1.4948786}
  {https://doi.org/10.1063/1.4948786} \BibitemShut {NoStop}%
\bibitem [{\citenamefont {Abedi}\ \emph
  {et~al.}(2013{\natexlab{b}})\citenamefont {Abedi}, \citenamefont {Agostini},
  \citenamefont {Suzuki},\ and\ \citenamefont {Gross}}]{AASG13}%
  \BibitemOpen
  \bibfield  {author} {\bibinfo {author} {\bibfnamefont {A.}~\bibnamefont
  {Abedi}}, \bibinfo {author} {\bibfnamefont {F.}~\bibnamefont {Agostini}},
  \bibinfo {author} {\bibfnamefont {Y.}~\bibnamefont {Suzuki}}, \ and\ \bibinfo
  {author} {\bibfnamefont {E.~K.~U.}\ \bibnamefont {Gross}},\ }\href@noop {}
  {\bibfield  {journal} {\bibinfo  {journal} {Phys. Rev. Lett.}\ }\textbf
  {\bibinfo {volume} {110}},\ \bibinfo {pages} {263001} (\bibinfo {year}
  {2013}{\natexlab{b}})}\BibitemShut {NoStop}%
\bibitem [{\citenamefont {Min}\ \emph {et~al.}(2014)\citenamefont {Min},
  \citenamefont {Abedi}, \citenamefont {Kim},\ and\ \citenamefont
  {Gross}}]{MAKG14}%
  \BibitemOpen
  \bibfield  {author} {\bibinfo {author} {\bibfnamefont {S.~K.}\ \bibnamefont
  {Min}}, \bibinfo {author} {\bibfnamefont {A.}~\bibnamefont {Abedi}}, \bibinfo
  {author} {\bibfnamefont {K.~S.}\ \bibnamefont {Kim}}, \ and\ \bibinfo
  {author} {\bibfnamefont {E.~K.~U.}\ \bibnamefont {Gross}},\ }\href@noop {}
  {\bibfield  {journal} {\bibinfo  {journal} {Phys. Rev. Lett.}\ }\textbf
  {\bibinfo {volume} {113}},\ \bibinfo {pages} {263004} (\bibinfo {year}
  {2014})}\BibitemShut {NoStop}%
\bibitem [{\citenamefont {Curchod}\ \emph {et~al.}(2016)\citenamefont
  {Curchod}, \citenamefont {Agostini},\ and\ \citenamefont {Gross}}]{CAG16}%
  \BibitemOpen
  \bibfield  {author} {\bibinfo {author} {\bibfnamefont {B.~F.~E.}\
  \bibnamefont {Curchod}}, \bibinfo {author} {\bibfnamefont {F.}~\bibnamefont
  {Agostini}}, \ and\ \bibinfo {author} {\bibfnamefont {E.~K.~U.}\ \bibnamefont
  {Gross}},\ }\href@noop {} {\bibfield  {journal} {\bibinfo  {journal} {J.
  Chem. Phys.}\ }\textbf {\bibinfo {volume} {145}},\ \bibinfo {pages} {034103}
  (\bibinfo {year} {2016})}\BibitemShut {NoStop}%
\bibitem [{\citenamefont {Fiedlschuster}\ \emph {et~al.}(2017)\citenamefont
  {Fiedlschuster}, \citenamefont {Handt}, \citenamefont {Gross},\ and\
  \citenamefont {Schmidt}}]{FHGS17}%
  \BibitemOpen
  \bibfield  {author} {\bibinfo {author} {\bibfnamefont {T.}~\bibnamefont
  {Fiedlschuster}}, \bibinfo {author} {\bibfnamefont {J.}~\bibnamefont
  {Handt}}, \bibinfo {author} {\bibfnamefont {E.~K.~U.}\ \bibnamefont {Gross}},
  \ and\ \bibinfo {author} {\bibfnamefont {R.}~\bibnamefont {Schmidt}},\ }\href
  {\doibase 10.1103/PhysRevA.95.063424} {\bibfield  {journal} {\bibinfo
  {journal} {Phys. Rev. A}\ }\textbf {\bibinfo {volume} {95}},\ \bibinfo
  {pages} {063424} (\bibinfo {year} {2017})}\BibitemShut {NoStop}%
\bibitem [{\citenamefont {Requist}\ \emph {et~al.}(2016)\citenamefont
  {Requist}, \citenamefont {Tandetzky},\ and\ \citenamefont {Gross}}]{RTG16}%
  \BibitemOpen
  \bibfield  {author} {\bibinfo {author} {\bibfnamefont {R.}~\bibnamefont
  {Requist}}, \bibinfo {author} {\bibfnamefont {F.}~\bibnamefont {Tandetzky}},
  \ and\ \bibinfo {author} {\bibfnamefont {E.~K.~U.}\ \bibnamefont {Gross}},\
  }\href@noop {} {\bibfield  {journal} {\bibinfo  {journal} {Phys. Rev. A}\
  }\textbf {\bibinfo {volume} {93}},\ \bibinfo {pages} {042108} (\bibinfo
  {year} {2016})}\BibitemShut {NoStop}%
\bibitem [{\citenamefont {Requist}\ \emph {et~al.}(2017)\citenamefont
  {Requist}, \citenamefont {Proetto},\ and\ \citenamefont {Gross}}]{RPG17}%
  \BibitemOpen
  \bibfield  {author} {\bibinfo {author} {\bibfnamefont {R.}~\bibnamefont
  {Requist}}, \bibinfo {author} {\bibfnamefont {C.~R.}\ \bibnamefont
  {Proetto}}, \ and\ \bibinfo {author} {\bibfnamefont {E.~K.~U.}\ \bibnamefont
  {Gross}},\ }\href@noop {} {\bibfield  {journal} {\bibinfo  {journal} {Phys.
  Rev. A}\ }\textbf {\bibinfo {volume} {96}},\ \bibinfo {pages} {062503}
  (\bibinfo {year} {2017})}\BibitemShut {NoStop}%
\bibitem [{\citenamefont {Curchod}\ and\ \citenamefont
  {Agostini}(2017)}]{CA17}%
  \BibitemOpen
  \bibfield  {author} {\bibinfo {author} {\bibfnamefont {B.~F.~E.}\
  \bibnamefont {Curchod}}\ and\ \bibinfo {author} {\bibfnamefont
  {F.}~\bibnamefont {Agostini}},\ }\href@noop {} {\bibfield  {journal}
  {\bibinfo  {journal} {J. Phys. Chem. Lett.}\ }\textbf {\bibinfo {volume}
  {8}},\ \bibinfo {pages} {831} (\bibinfo {year} {2017})}\BibitemShut {NoStop}%
\bibitem [{\citenamefont {Chiang}\ \emph {et~al.}(2014)\citenamefont {Chiang},
  \citenamefont {Klaiman}, \citenamefont {Otto},\ and\ \citenamefont
  {Cederbaum}}]{CKOC14}%
  \BibitemOpen
  \bibfield  {author} {\bibinfo {author} {\bibfnamefont {Y.-C.}\ \bibnamefont
  {Chiang}}, \bibinfo {author} {\bibfnamefont {S.}~\bibnamefont {Klaiman}},
  \bibinfo {author} {\bibfnamefont {F.}~\bibnamefont {Otto}}, \ and\ \bibinfo
  {author} {\bibfnamefont {L.~S.}\ \bibnamefont {Cederbaum}},\ }\href@noop {}
  {\bibfield  {journal} {\bibinfo  {journal} {J. Chem. Phys.}\ }\textbf
  {\bibinfo {volume} {140}},\ \bibinfo {pages} {054104} (\bibinfo {year}
  {2014})}\BibitemShut {NoStop}%
\bibitem [{\citenamefont {Lefebvre}(2015{\natexlab{a}})}]{L15}%
  \BibitemOpen
  \bibfield  {author} {\bibinfo {author} {\bibfnamefont {R.}~\bibnamefont
  {Lefebvre}},\ }\href@noop {} {\bibfield  {journal} {\bibinfo  {journal} {J.
  Chem. Phys.}\ }\textbf {\bibinfo {volume} {142}},\ \bibinfo {pages} {074106}
  (\bibinfo {year} {2015}{\natexlab{a}})}\BibitemShut {NoStop}%
\bibitem [{\citenamefont {Lefebvre}(2015{\natexlab{b}})}]{L15b}%
  \BibitemOpen
  \bibfield  {author} {\bibinfo {author} {\bibfnamefont {R.}~\bibnamefont
  {Lefebvre}},\ }\href@noop {} {\bibfield  {journal} {\bibinfo  {journal} {J.
  Chem. Phys.}\ }\textbf {\bibinfo {volume} {142}},\ \bibinfo {pages} {214105}
  (\bibinfo {year} {2015}{\natexlab{b}})}\BibitemShut {NoStop}%
\bibitem [{\citenamefont {Agostini}\ and\ \citenamefont
  {Curchod}(2018)}]{Curchod_EPJB2018}%
  \BibitemOpen
  \bibfield  {author} {\bibinfo {author} {\bibfnamefont {F.}~\bibnamefont
  {Agostini}}\ and\ \bibinfo {author} {\bibfnamefont {B.~F.~E.}\ \bibnamefont
  {Curchod}},\ }\href@noop {} {\bibfield  {journal} {\bibinfo  {journal} {Euro.
  Phys. J. B}\ }\textbf {\bibinfo {volume} {submitted}} (\bibinfo {year}
  {2018})}\BibitemShut {NoStop}%
\bibitem [{\citenamefont {Scherrer}\ \emph {et~al.}(2017)\citenamefont
  {Scherrer}, \citenamefont {Agostini}, \citenamefont {Sebastiani},
  \citenamefont {Gross},\ and\ \citenamefont {Vuilleumier}}]{SASGV17}%
  \BibitemOpen
  \bibfield  {author} {\bibinfo {author} {\bibfnamefont {A.}~\bibnamefont
  {Scherrer}}, \bibinfo {author} {\bibfnamefont {F.}~\bibnamefont {Agostini}},
  \bibinfo {author} {\bibfnamefont {D.}~\bibnamefont {Sebastiani}}, \bibinfo
  {author} {\bibfnamefont {E.~K.~U.}\ \bibnamefont {Gross}}, \ and\ \bibinfo
  {author} {\bibfnamefont {R.}~\bibnamefont {Vuilleumier}},\ }\href@noop {}
  {\bibfield  {journal} {\bibinfo  {journal} {Phys. Rev. X}\ }\textbf {\bibinfo
  {volume} {7}},\ \bibinfo {pages} {031035} (\bibinfo {year}
  {2017})}\BibitemShut {NoStop}%
\bibitem [{\citenamefont {Scherrer}\ \emph {et~al.}(2015)\citenamefont
  {Scherrer}, \citenamefont {Agostini}, \citenamefont {Sebastiani},
  \citenamefont {Gross},\ and\ \citenamefont {Vuilleumier}}]{SASGV15}%
  \BibitemOpen
  \bibfield  {author} {\bibinfo {author} {\bibfnamefont {A.}~\bibnamefont
  {Scherrer}}, \bibinfo {author} {\bibfnamefont {F.}~\bibnamefont {Agostini}},
  \bibinfo {author} {\bibfnamefont {D.}~\bibnamefont {Sebastiani}}, \bibinfo
  {author} {\bibfnamefont {E.~K.~U.}\ \bibnamefont {Gross}}, \ and\ \bibinfo
  {author} {\bibfnamefont {R.}~\bibnamefont {Vuilleumier}},\ }\href@noop {}
  {\bibfield  {journal} {\bibinfo  {journal} {J. Chem. Phys.}\ }\textbf
  {\bibinfo {volume} {143}},\ \bibinfo {pages} {074106} (\bibinfo {year}
  {2015})}\BibitemShut {NoStop}%
\bibitem [{\citenamefont {Eich}\ and\ \citenamefont {Agostini}(2016)}]{EA16}%
  \BibitemOpen
  \bibfield  {author} {\bibinfo {author} {\bibfnamefont {F.~G.}\ \bibnamefont
  {Eich}}\ and\ \bibinfo {author} {\bibfnamefont {F.}~\bibnamefont
  {Agostini}},\ }\href@noop {} {\bibfield  {journal} {\bibinfo  {journal} {J.
  Chem. Phys.}\ }\textbf {\bibinfo {volume} {145}},\ \bibinfo {pages} {054110}
  (\bibinfo {year} {2016})}\BibitemShut {NoStop}%
\bibitem [{\citenamefont {Schild}\ \emph {et~al.}(2016)\citenamefont {Schild},
  \citenamefont {Agostini},\ and\ \citenamefont {Gross}}]{SAG16}%
  \BibitemOpen
  \bibfield  {author} {\bibinfo {author} {\bibfnamefont {A.}~\bibnamefont
  {Schild}}, \bibinfo {author} {\bibfnamefont {F.}~\bibnamefont {Agostini}}, \
  and\ \bibinfo {author} {\bibfnamefont {E.~K.~U.}\ \bibnamefont {Gross}},\
  }\href@noop {} {\bibfield  {journal} {\bibinfo  {journal} {J. Phys. Chem. A}\
  }\textbf {\bibinfo {volume} {120}},\ \bibinfo {pages} {3316} (\bibinfo {year}
  {2016})}\BibitemShut {NoStop}%
\bibitem [{\citenamefont {Requist}\ and\ \citenamefont {Gross}(2016)}]{RG16}%
  \BibitemOpen
  \bibfield  {author} {\bibinfo {author} {\bibfnamefont {R.}~\bibnamefont
  {Requist}}\ and\ \bibinfo {author} {\bibfnamefont {E.~K.~U.}\ \bibnamefont
  {Gross}},\ }\href@noop {} {\bibfield  {journal} {\bibinfo  {journal} {Phys.
  Rev. Lett.}\ }\textbf {\bibinfo {volume} {117}},\ \bibinfo {pages} {193001}
  (\bibinfo {year} {2016})}\BibitemShut {NoStop}%
\bibitem [{\citenamefont {Li}\ \emph {et~al.}(2018)\citenamefont {Li},
  \citenamefont {Requist},\ and\ \citenamefont {Gross}}]{LRG18}%
  \BibitemOpen
  \bibfield  {author} {\bibinfo {author} {\bibfnamefont {C.}~\bibnamefont
  {Li}}, \bibinfo {author} {\bibfnamefont {R.}~\bibnamefont {Requist}}, \ and\
  \bibinfo {author} {\bibfnamefont {E.~K.~U.}\ \bibnamefont {Gross}},\ }\href
  {\doibase 10.1063/1.5011663} {\bibfield  {journal} {\bibinfo  {journal} {The
  Journal of Chemical Physics}\ }\textbf {\bibinfo {volume} {148}},\ \bibinfo
  {pages} {084110} (\bibinfo {year} {2018})},\ \Eprint
  {http://arxiv.org/abs/https://doi.org/10.1063/1.5011663}
  {https://doi.org/10.1063/1.5011663} \BibitemShut {NoStop}%
\bibitem [{\citenamefont {Cederbaum}(2015)}]{C15}%
  \BibitemOpen
  \bibfield  {author} {\bibinfo {author} {\bibfnamefont {L.~S.}\ \bibnamefont
  {Cederbaum}},\ }\href@noop {} {\bibfield  {journal} {\bibinfo  {journal}
  {Chem. Phys.}\ }\textbf {\bibinfo {volume} {457}},\ \bibinfo {pages} {129}
  (\bibinfo {year} {2015})}\BibitemShut {NoStop}%
\bibitem [{\citenamefont {Schild}\ and\ \citenamefont {Gross}(2017)}]{SG17}%
  \BibitemOpen
  \bibfield  {author} {\bibinfo {author} {\bibfnamefont {A.}~\bibnamefont
  {Schild}}\ and\ \bibinfo {author} {\bibfnamefont {E.~K.~U.}\ \bibnamefont
  {Gross}},\ }\href {\doibase 10.1103/PhysRevLett.118.163202} {\bibfield
  {journal} {\bibinfo  {journal} {Phys. Rev. Lett.}\ }\textbf {\bibinfo
  {volume} {118}},\ \bibinfo {pages} {163202} (\bibinfo {year}
  {2017})}\BibitemShut {NoStop}%
\bibitem [{\citenamefont {Hoffmann}\ \emph {et~al.}(2018)\citenamefont
  {Hoffmann}, \citenamefont {Appel}, \citenamefont {Rubio},\ and\ \citenamefont
  {Maitra}}]{HARM18}%
  \BibitemOpen
  \bibfield  {author} {\bibinfo {author} {\bibfnamefont {N.~M.}\ \bibnamefont
  {Hoffmann}}, \bibinfo {author} {\bibfnamefont {H.}~\bibnamefont {Appel}},
  \bibinfo {author} {\bibfnamefont {A.}~\bibnamefont {Rubio}}, \ and\ \bibinfo
  {author} {\bibfnamefont {N.}~\bibnamefont {Maitra}},\ }\href@noop {}
  {\bibfield  {journal} {\bibinfo  {journal} {arXiv:1803.02020}\ } (\bibinfo
  {year} {2018})}\BibitemShut {NoStop}%
\bibitem [{\citenamefont {Agostini}\ \emph
  {et~al.}(2015{\natexlab{a}})\citenamefont {Agostini}, \citenamefont {Min},\
  and\ \citenamefont {Gross}}]{AMG15}%
  \BibitemOpen
  \bibfield  {author} {\bibinfo {author} {\bibfnamefont {F.}~\bibnamefont
  {Agostini}}, \bibinfo {author} {\bibfnamefont {S.~K.}\ \bibnamefont {Min}}, \
  and\ \bibinfo {author} {\bibfnamefont {E.~K.~U.}\ \bibnamefont {Gross}},\
  }\href@noop {} {\bibfield  {journal} {\bibinfo  {journal} {Ann. Phys.}\
  }\textbf {\bibinfo {volume} {527}},\ \bibinfo {pages} {546} (\bibinfo {year}
  {2015}{\natexlab{a}})}\BibitemShut {NoStop}%
\bibitem [{\citenamefont {Agostini}\ \emph {et~al.}(2014)\citenamefont
  {Agostini}, \citenamefont {Abedi},\ and\ \citenamefont {Gross}}]{AAG14}%
  \BibitemOpen
  \bibfield  {author} {\bibinfo {author} {\bibfnamefont {F.}~\bibnamefont
  {Agostini}}, \bibinfo {author} {\bibfnamefont {A.}~\bibnamefont {Abedi}}, \
  and\ \bibinfo {author} {\bibfnamefont {E.~K.~U.}\ \bibnamefont {Gross}},\
  }\href@noop {} {\bibfield  {journal} {\bibinfo  {journal} {J. Chem. Phys.}\
  }\textbf {\bibinfo {volume} {141}},\ \bibinfo {pages} {214101} (\bibinfo
  {year} {2014})}\BibitemShut {NoStop}%
\bibitem [{\citenamefont {Abedi}\ \emph {et~al.}(2014)\citenamefont {Abedi},
  \citenamefont {Agostini},\ and\ \citenamefont {Gross}}]{AAG14b}%
  \BibitemOpen
  \bibfield  {author} {\bibinfo {author} {\bibfnamefont {A.}~\bibnamefont
  {Abedi}}, \bibinfo {author} {\bibfnamefont {F.}~\bibnamefont {Agostini}}, \
  and\ \bibinfo {author} {\bibfnamefont {E.~K.~U.}\ \bibnamefont {Gross}},\
  }\href@noop {} {\bibfield  {journal} {\bibinfo  {journal} {Europhys. Lett.}\
  }\textbf {\bibinfo {volume} {106}},\ \bibinfo {pages} {33001} (\bibinfo
  {year} {2014})}\BibitemShut {NoStop}%
\bibitem [{\citenamefont {Agostini}\ \emph
  {et~al.}(2015{\natexlab{b}})\citenamefont {Agostini}, \citenamefont {Abedi},
  \citenamefont {Suzuki}, \citenamefont {Min}, \citenamefont {Maitra},\ and\
  \citenamefont {Gross}}]{AASMMG15}%
  \BibitemOpen
  \bibfield  {author} {\bibinfo {author} {\bibfnamefont {F.}~\bibnamefont
  {Agostini}}, \bibinfo {author} {\bibfnamefont {A.}~\bibnamefont {Abedi}},
  \bibinfo {author} {\bibfnamefont {Y.}~\bibnamefont {Suzuki}}, \bibinfo
  {author} {\bibfnamefont {S.~K.}\ \bibnamefont {Min}}, \bibinfo {author}
  {\bibfnamefont {N.~T.}\ \bibnamefont {Maitra}}, \ and\ \bibinfo {author}
  {\bibfnamefont {E.~K.~U.}\ \bibnamefont {Gross}},\ }\href@noop {} {\bibfield
  {journal} {\bibinfo  {journal} {J. Chem. Phys.}\ }\textbf {\bibinfo {volume}
  {142}},\ \bibinfo {pages} {084303} (\bibinfo {year}
  {2015}{\natexlab{b}})}\BibitemShut {NoStop}%
\bibitem [{\citenamefont {Suzuki}\ \emph {et~al.}(2015)\citenamefont {Suzuki},
  \citenamefont {Abedi}, \citenamefont {Maitra},\ and\ \citenamefont
  {Gross}}]{SAMG15}%
  \BibitemOpen
  \bibfield  {author} {\bibinfo {author} {\bibfnamefont {Y.}~\bibnamefont
  {Suzuki}}, \bibinfo {author} {\bibfnamefont {A.}~\bibnamefont {Abedi}},
  \bibinfo {author} {\bibfnamefont {N.~T.}\ \bibnamefont {Maitra}}, \ and\
  \bibinfo {author} {\bibfnamefont {E.~K.~U.}\ \bibnamefont {Gross}},\
  }\href@noop {} {\bibfield  {journal} {\bibinfo  {journal} {Phys. Chem. Chem.
  Phys.}\ }\textbf {\bibinfo {volume} {17}},\ \bibinfo {pages} {29271}
  (\bibinfo {year} {2015})}\BibitemShut {NoStop}%
\bibitem [{\citenamefont {Suzuki}\ and\ \citenamefont {Watanabe}(2016)}]{SW16}%
  \BibitemOpen
  \bibfield  {author} {\bibinfo {author} {\bibfnamefont {Y.}~\bibnamefont
  {Suzuki}}\ and\ \bibinfo {author} {\bibfnamefont {K.}~\bibnamefont
  {Watanabe}},\ }\href@noop {} {\bibfield  {journal} {\bibinfo  {journal}
  {Phys. Rev. A}\ }\textbf {\bibinfo {volume} {94}},\ \bibinfo {pages} {032517}
  (\bibinfo {year} {2016})}\BibitemShut {NoStop}%
\bibitem [{\citenamefont {Ha}\ \emph {et~al.}(2018)\citenamefont {Ha},
  \citenamefont {Lee},\ and\ \citenamefont {Min}}]{HLM18}%
  \BibitemOpen
  \bibfield  {author} {\bibinfo {author} {\bibfnamefont {J.-K.}\ \bibnamefont
  {Ha}}, \bibinfo {author} {\bibfnamefont {I.~S.}\ \bibnamefont {Lee}}, \ and\
  \bibinfo {author} {\bibfnamefont {S.~K.}\ \bibnamefont {Min}},\ }\href
  {\doibase 10.1021/acs.jpclett.8b00060} {\bibfield  {journal} {\bibinfo
  {journal} {The Journal of Physical Chemistry Letters}\ }\textbf {\bibinfo
  {volume} {9}},\ \bibinfo {pages} {1097} (\bibinfo {year} {2018})},\ \bibinfo
  {note} {pMID: 29439572},\ \Eprint
  {http://arxiv.org/abs/https://doi.org/10.1021/acs.jpclett.8b00060}
  {https://doi.org/10.1021/acs.jpclett.8b00060} \BibitemShut {NoStop}%
\bibitem [{\citenamefont {Gu}\ and\ \citenamefont {Franco}(2017)}]{GF17}%
  \BibitemOpen
  \bibfield  {author} {\bibinfo {author} {\bibfnamefont {B.}~\bibnamefont
  {Gu}}\ and\ \bibinfo {author} {\bibfnamefont {I.}~\bibnamefont {Franco}},\
  }\href {\doibase 10.1063/1.4983495} {\bibfield  {journal} {\bibinfo
  {journal} {J. Chem. Phys.}\ }\textbf {\bibinfo {volume} {146}},\ \bibinfo
  {pages} {194104} (\bibinfo {year} {2017})},\ \Eprint
  {http://arxiv.org/abs/https://doi.org/10.1063/1.4983495}
  {https://doi.org/10.1063/1.4983495} \BibitemShut {NoStop}%
\bibitem [{\citenamefont {Curchod}\ \emph {et~al.}(2018)\citenamefont
  {Curchod}, \citenamefont {Agostini},\ and\ \citenamefont
  {Tavernelli}}]{Tavernelli_EPJB2018}%
  \BibitemOpen
  \bibfield  {author} {\bibinfo {author} {\bibfnamefont {B.~F.~E.}\
  \bibnamefont {Curchod}}, \bibinfo {author} {\bibfnamefont {F.}~\bibnamefont
  {Agostini}}, \ and\ \bibinfo {author} {\bibfnamefont {I.}~\bibnamefont
  {Tavernelli}},\ }\href@noop {} {\bibfield  {journal} {\bibinfo  {journal}
  {Euro. Phys. J. B}\ }\textbf {\bibinfo {volume} {submitted}} (\bibinfo {year}
  {2018})}\BibitemShut {NoStop}%
\bibitem [{\citenamefont {Agostini}\ \emph
  {et~al.}(2018{\natexlab{a}})\citenamefont {Agostini}, \citenamefont
  {Curchod}, \citenamefont {Vuilleumier}, \citenamefont {Tavernelli},\ and\
  \citenamefont {Gross}}]{Gross_bookTDDFT2018}%
  \BibitemOpen
  \bibfield  {author} {\bibinfo {author} {\bibfnamefont {F.}~\bibnamefont
  {Agostini}}, \bibinfo {author} {\bibfnamefont {B.~F.~E.}\ \bibnamefont
  {Curchod}}, \bibinfo {author} {\bibfnamefont {R.}~\bibnamefont
  {Vuilleumier}}, \bibinfo {author} {\bibfnamefont {I.}~\bibnamefont
  {Tavernelli}}, \ and\ \bibinfo {author} {\bibfnamefont {E.~K.~U.}\
  \bibnamefont {Gross}},\ }in\ \href@noop {} {\emph {\bibinfo {booktitle}
  {Handbook of Materials Modeling. {Vol. 1: T}heory and Modeling}}},\ \bibinfo
  {editor} {edited by\ \bibinfo {editor} {\bibfnamefont {W.}~\bibnamefont
  {Andreoni}}\ and\ \bibinfo {editor} {\bibfnamefont {S.}~\bibnamefont {Yip}}}\
  (\bibinfo  {publisher} {Springer},\ \bibinfo {year} {2018})\ p.\ \bibinfo
  {pages} {accepted}\BibitemShut {NoStop}%
\bibitem [{Note1()}]{Note1}%
  \BibitemOpen
  \bibinfo {note} {In the cases studied here, all surfaces have zero slope in
  the region where the nuclear wavepacket is initiated, thus its speed and
  those of the trajectories are simply determined by the initial average
  momentum.}\BibitemShut {Stop}%
\bibitem [{\citenamefont {Agostini}(2018)}]{Agostini_EPJB2018}%
  \BibitemOpen
  \bibfield  {author} {\bibinfo {author} {\bibfnamefont {F.}~\bibnamefont
  {Agostini}},\ }\href@noop {} {\bibfield  {journal} {\bibinfo  {journal}
  {Euro. Phys. J. B}\ }\textbf {\bibinfo {volume} {submitted}} (\bibinfo {year}
  {2018})}\BibitemShut {NoStop}%
\bibitem [{\citenamefont {Agostini}\ \emph
  {et~al.}(2018{\natexlab{b}})\citenamefont {Agostini}, \citenamefont
  {Tavernelli},\ and\ \citenamefont {Ciccotti}}]{Ciccotti_EPJB2018}%
  \BibitemOpen
  \bibfield  {author} {\bibinfo {author} {\bibfnamefont {F.}~\bibnamefont
  {Agostini}}, \bibinfo {author} {\bibfnamefont {I.}~\bibnamefont
  {Tavernelli}}, \ and\ \bibinfo {author} {\bibfnamefont {G.}~\bibnamefont
  {Ciccotti}},\ }\href@noop {} {\bibfield  {journal} {\bibinfo  {journal}
  {Euro. Phys. J. B}\ }\textbf {\bibinfo {volume} {submitted}} (\bibinfo {year}
  {2018}{\natexlab{b}})}\BibitemShut {NoStop}%
\bibitem [{Note2()}]{Note2}%
  \BibitemOpen
  \bibinfo {note} {One might imagine the coupled-trajectory term in the nuclear
  equation can allow the classical momentum $P^{(I)}(t) $ to split due to the
  structure of $\protect \mathcal {Q}^{(I)}$, which could possibly lead to
  branching of the electronic populations. However, this did not arise in any
  of the examples we considered.}\BibitemShut {Stop}%
\bibitem [{\citenamefont {Schwartz}\ \emph {et~al.}(1996)\citenamefont
  {Schwartz}, \citenamefont {Bittner}, \citenamefont {Prezhdo},\ and\
  \citenamefont {Rossky}}]{SBPR96}%
  \BibitemOpen
  \bibfield  {author} {\bibinfo {author} {\bibfnamefont {B.~J.}\ \bibnamefont
  {Schwartz}}, \bibinfo {author} {\bibfnamefont {E.~R.}\ \bibnamefont
  {Bittner}}, \bibinfo {author} {\bibfnamefont {O.~V.}\ \bibnamefont
  {Prezhdo}}, \ and\ \bibinfo {author} {\bibfnamefont {P.~J.}\ \bibnamefont
  {Rossky}},\ }\href {\doibase 10.1063/1.471326} {\bibfield  {journal}
  {\bibinfo  {journal} {The Journal of Chemical Physics}\ }\textbf {\bibinfo
  {volume} {104}},\ \bibinfo {pages} {5942} (\bibinfo {year} {1996})},\ \Eprint
  {http://arxiv.org/abs/https://doi.org/10.1063/1.471326}
  {https://doi.org/10.1063/1.471326} \BibitemShut {NoStop}%
\bibitem [{\citenamefont {Prezhdo}\ and\ \citenamefont {Rossky}(1997)}]{PR97}%
  \BibitemOpen
  \bibfield  {author} {\bibinfo {author} {\bibfnamefont {O.~V.}\ \bibnamefont
  {Prezhdo}}\ and\ \bibinfo {author} {\bibfnamefont {P.~J.}\ \bibnamefont
  {Rossky}},\ }\href@noop {} {\bibfield  {journal} {\bibinfo  {journal} {J.
  Chem. Phys.}\ }\textbf {\bibinfo {volume} {107}},\ \bibinfo {pages} {825}
  (\bibinfo {year} {1997})}\BibitemShut {NoStop}%
\bibitem [{\citenamefont {Wang}\ \emph
  {et~al.}(2016{\natexlab{b}})\citenamefont {Wang}, \citenamefont {Akimov},\
  and\ \citenamefont {Prezhdo}}]{WAP16}%
  \BibitemOpen
  \bibfield  {author} {\bibinfo {author} {\bibfnamefont {L.}~\bibnamefont
  {Wang}}, \bibinfo {author} {\bibfnamefont {A.}~\bibnamefont {Akimov}}, \ and\
  \bibinfo {author} {\bibfnamefont {O.~V.}\ \bibnamefont {Prezhdo}},\
  }\href@noop {} {\bibfield  {journal} {\bibinfo  {journal} {J. Phys. Chem.
  Lett.}\ }\textbf {\bibinfo {volume} {7}},\ \bibinfo {pages} {2100} (\bibinfo
  {year} {2016}{\natexlab{b}})}\BibitemShut {NoStop}%
\bibitem [{\citenamefont {Zhu}\ \emph {et~al.}(2004)\citenamefont {Zhu},
  \citenamefont {Nangia}, \citenamefont {Jasper},\ and\ \citenamefont
  {Truhlar}}]{ZNJT04}%
  \BibitemOpen
  \bibfield  {author} {\bibinfo {author} {\bibfnamefont {C.}~\bibnamefont
  {Zhu}}, \bibinfo {author} {\bibfnamefont {S.}~\bibnamefont {Nangia}},
  \bibinfo {author} {\bibfnamefont {A.~W.}\ \bibnamefont {Jasper}}, \ and\
  \bibinfo {author} {\bibfnamefont {D.~G.}\ \bibnamefont {Truhlar}},\
  }\href@noop {} {\bibfield  {journal} {\bibinfo  {journal} {J. Chem. Phys.}\
  }\textbf {\bibinfo {volume} {121}},\ \bibinfo {pages} {7658} (\bibinfo {year}
  {2004})}\BibitemShut {NoStop}%
\bibitem [{\citenamefont {Granucci}\ and\ \citenamefont
  {Persico}(2007)}]{GP07}%
  \BibitemOpen
  \bibfield  {author} {\bibinfo {author} {\bibfnamefont {G.}~\bibnamefont
  {Granucci}}\ and\ \bibinfo {author} {\bibfnamefont {M.}~\bibnamefont
  {Persico}},\ }\href@noop {} {\bibfield  {journal} {\bibinfo  {journal} {J.
  Chem. Phys.}\ }\textbf {\bibinfo {volume} {126}},\ \bibinfo {pages} {134114}
  (\bibinfo {year} {2007})}\BibitemShut {NoStop}%
\bibitem [{\citenamefont {Martens}(2016)}]{M16b}%
  \BibitemOpen
  \bibfield  {author} {\bibinfo {author} {\bibfnamefont {C.~C.}\ \bibnamefont
  {Martens}},\ }\href {\doibase 10.1021/acs.jpclett.6b01186} {\bibfield
  {journal} {\bibinfo  {journal} {The Journal of Physical Chemistry Letters}\
  }\textbf {\bibinfo {volume} {7}},\ \bibinfo {pages} {2610} (\bibinfo {year}
  {2016})},\ \bibinfo {note} {pMID: 27345103},\ \Eprint
  {http://arxiv.org/abs/https://doi.org/10.1021/acs.jpclett.6b01186}
  {https://doi.org/10.1021/acs.jpclett.6b01186} \BibitemShut {NoStop}%
\bibitem [{\citenamefont {Gao}\ and\ \citenamefont {Thiel}(2017)}]{GT17}%
  \BibitemOpen
  \bibfield  {author} {\bibinfo {author} {\bibfnamefont {X.}~\bibnamefont
  {Gao}}\ and\ \bibinfo {author} {\bibfnamefont {W.}~\bibnamefont {Thiel}},\
  }\href {\doibase 10.1103/PhysRevE.95.013308} {\bibfield  {journal} {\bibinfo
  {journal} {Phys. Rev. E}\ }\textbf {\bibinfo {volume} {95}},\ \bibinfo
  {pages} {013308} (\bibinfo {year} {2017})}\BibitemShut {NoStop}%
\bibitem [{\citenamefont {Kapral}\ and\ \citenamefont {Ciccotti}(1999)}]{KC99}%
  \BibitemOpen
  \bibfield  {author} {\bibinfo {author} {\bibfnamefont {R.}~\bibnamefont
  {Kapral}}\ and\ \bibinfo {author} {\bibfnamefont {G.}~\bibnamefont
  {Ciccotti}},\ }\href@noop {} {\bibfield  {journal} {\bibinfo  {journal} {J.
  Chem.Phys.}\ }\textbf {\bibinfo {volume} {110}},\ \bibinfo {pages} {8916}
  (\bibinfo {year} {1999})}\BibitemShut {NoStop}%
\bibitem [{\citenamefont {Feit}\ \emph {et~al.}(1982)\citenamefont {Feit},
  \citenamefont {{F. A. Fleck Jr.}},\ and\ \citenamefont {Steiger}}]{spo}%
  \BibitemOpen
  \bibfield  {author} {\bibinfo {author} {\bibfnamefont {M.~D.}\ \bibnamefont
  {Feit}}, \bibinfo {author} {\bibnamefont {{F. A. Fleck Jr.}}}, \ and\
  \bibinfo {author} {\bibfnamefont {A.}~\bibnamefont {Steiger}},\ }\href@noop
  {} {\bibfield  {journal} {\bibinfo  {journal} {J. Comput. Phys.}\ }\textbf
  {\bibinfo {volume} {47}},\ \bibinfo {pages} {412} (\bibinfo {year}
  {1982})}\BibitemShut {NoStop}%
\end{thebibliography}%

\end{document}